    \documentclass[
        fleqn,
        usenatbib,
    ]{mnras}
    \usepackage{
        amsmath,
        amssymb,
        newtxtext,
        newtxmath,
        ae, aecompl,
        graphicx,
        booktabs,
        xcolor,
    }

    \newcommand*{\rd}[2]{\frac{\mathrm{d}#1}{\mathrm{d}#2}}

    \newcommand*{\rdil}[2]{\mathrm{d}#1/\mathrm{d}#2}
    \newcommand*{\at}[1]{\left.#1\right|}
    \newcommand*{\abs}[1]{\left|#1\right|}
    \newcommand*{\ev}[1]{\left\langle#1\right\rangle}
    \newcommand*{\p}[1]{\left(#1\right)}
    \newcommand*{\s}[1]{\left[#1\right]}
    \newcommand*{\z}[1]{\left\{#1\right\}}
    \newcommand*{\bm}[1]{\mathbf{#1}}
    \newcommand*{\uv}[1]{\hat{\mathbf{#1}}}
    \newcommand*{\md}[0]{\mathrm{d}}

\title[Mass Ratio Distribution]{The Mass Ratio Distribution of
Tertiary Induced Binary Black Hole Mergers}
\author[Y. Su et\ al.]{
Yubo Su,$^1$\thanks{E-mail: yubosu@astro.cornell.edu},
Bin Liu,$^{1,2}$,
Dong Lai$^{1,3}$
\\
$^1$ Cornell Center for Astrophysics and Planetary Science, Department of
Astronomy, Cornell University, Ithaca, NY 14853, USA\\
$^2$ Niels Bohr International Academy, Niels Bohr Institute, Blegdamsvej 17,
2100 Copenhagen, Denmark\\
$^3$ Tsung-Dao Lee Institute \& School of Physics and Astronomy, Shanghai Jiao
Tong University, 200240 Shanghai, China
}

\date{Accepted June 02, 2021\@. Received March 02, 2021\@; in original form
March 02, 2021}

\pubyear{2021}

\begin{document}\label{firstpage}
\pagerange{\pageref{firstpage}--\pageref{lastpage}}
\maketitle

\begin{abstract}
    Many proposed scenarios for black hole (BH) mergers involve a tertiary
    companion that induces von Zeipel-Lidov-Kozai (ZLK) eccentricity cycles in
    the inner binary. An attractive feature of such mechanisms is the enhanced
    merger probability when the octupole-order effects, also known as the
    eccentric Kozai mechanism, are important. This can be the case when the
    tertiary is of comparable mass to the binary components. Since the octupole
    strength [$\propto (1-q)/(1+q)$] increases with decreasing binary mass ratio
    $q$, such ZLK-induced mergers favor binaries with smaller mass ratios. We
    use a combination of numerical and analytical approaches to fully
    characterize the octupole-enhanced binary BH mergers and provide
    semi-analytical criteria for efficiently calculating the
    strength of this enhancement. We show that for hierarchical triples with
    semi-major axis ratio $a/a_{\rm out}\gtrsim 0.01$--$0.02$, the binary merger
    fraction can increase by a large factor (up to $\sim 20$) as $q$ decreases
    from unity to $0.2$. The resulting mass ratio distribution for merging
    binary BHs produced in this scenario is in tension with the observed
    distribution obtained by the LIGO/VIRGO collaboration, although significant
    uncertainties remain about the initial distribution of binary BH masses and
    mass ratios.
\end{abstract}


\section{Introduction}\label{s:intro}

The $50$ or so black hole (BH) binary mergers detected by the LIGO/VIRGO
collaboration to date \citep{LIGOO3a} continue to motivate theoretical studies of their
formation channels. These range from the traditional isolated binary evolution,
in which mass transfer and friction in the common envelope phase cause the
binary orbit to decay sufficiently that it subsequently merges via emission of
gravitational waves (GWs) \citep[e.g.,][]{lipunov1997black,
lipunov2017first, podsiadlowski2003formation, belczynski2010effect,
belczynski2016first, dominik2012double, dominik2013double, dominik2015double},
to various flavors of dynamical formation channels that involve either strong
gravitational scatterings in dense clusters \citep[e.g.,][]{zwart1999black,
o2006binary, miller2009mergers, banerjee2010stellar, downing2010compact,
ziosi2014dynamics, rodriguez2015binary, samsing2017assembly, samsing2018black,
rodriguez2018post, gondan2018eccentric} or mergers in isolated triple and
quadruple systems induced by distant companions \citep[e.g.,][]{miller2002four,
wen2003eccentricity, antonini2012secular, antonini2017binary, silsbee2017lidov,
LL17, LL18, randall2018induced, randall2018analytical, hoang2018black,
fragione2019, fragione2019loeb, bin_misc5, LL19, LLW_apjl, bin_misc2,
bin_misc1}.

Given the large number of merger events to be detected in the coming years, it
is important to search for observational signatures to distinguish various BH
binary formation channels. The masses of merging BHs obviously carry important
information. The recent detection of BH binary systems with component masses in
the mass gap (such in GW190521) suggests that some kinds of ``hierarchical
mergers'' may be needed to explain these exceptional events (\citealp{190521};
see \citealp{bin_misc1} for examples of such ``hierarchical mergers'' in stellar
multiples). Another possible indicator is merger eccentricity: previous
studies find that dynamical binary-single interactions in dense clusters
\citep[e.g.,][]{samsing2017assembly, rodriguez2018post, samsing2018black,
fragione2019bromberg} or in galactic triples \citep{silsbee2017lidov,
antonini2017binary, fragione2019loeb, LL19} may lead to BH binaries that enter
the LIGO band with modest eccentricities. The third possible indicator is the
spin-orbit misalignment of the binary. In particular, the mass-weighted
projection of the BH spins,
\begin{equation}
    \chi_{\rm eff} = \frac{m_1 \boldsymbol{\chi}_1 + m_2\boldsymbol{\chi}_2}
    {m_1 + m_2} \cdot \uv{L},
\end{equation}
can be measured through the binary inspiral waveform [here, $m_{1,2}$ is the BH
mass, $\boldsymbol{\chi}_{1,2} = c\boldsymbol{S}_{1,2} / (Gm_{1,2}^2)$ is the
dimensionless BH spin, and $\uv{L}$ is the unit orbital angular momentum vector
of the binary]. Different formation histories yield different distributions of
$\chi_{\rm eff}$ \citep{LL17, LL18, antonini2018precessional, rodriguez2018post,
gerosa2018, LL19, su2020spin}.

\begin{figure}
    \centering
    \includegraphics[width=\columnwidth]{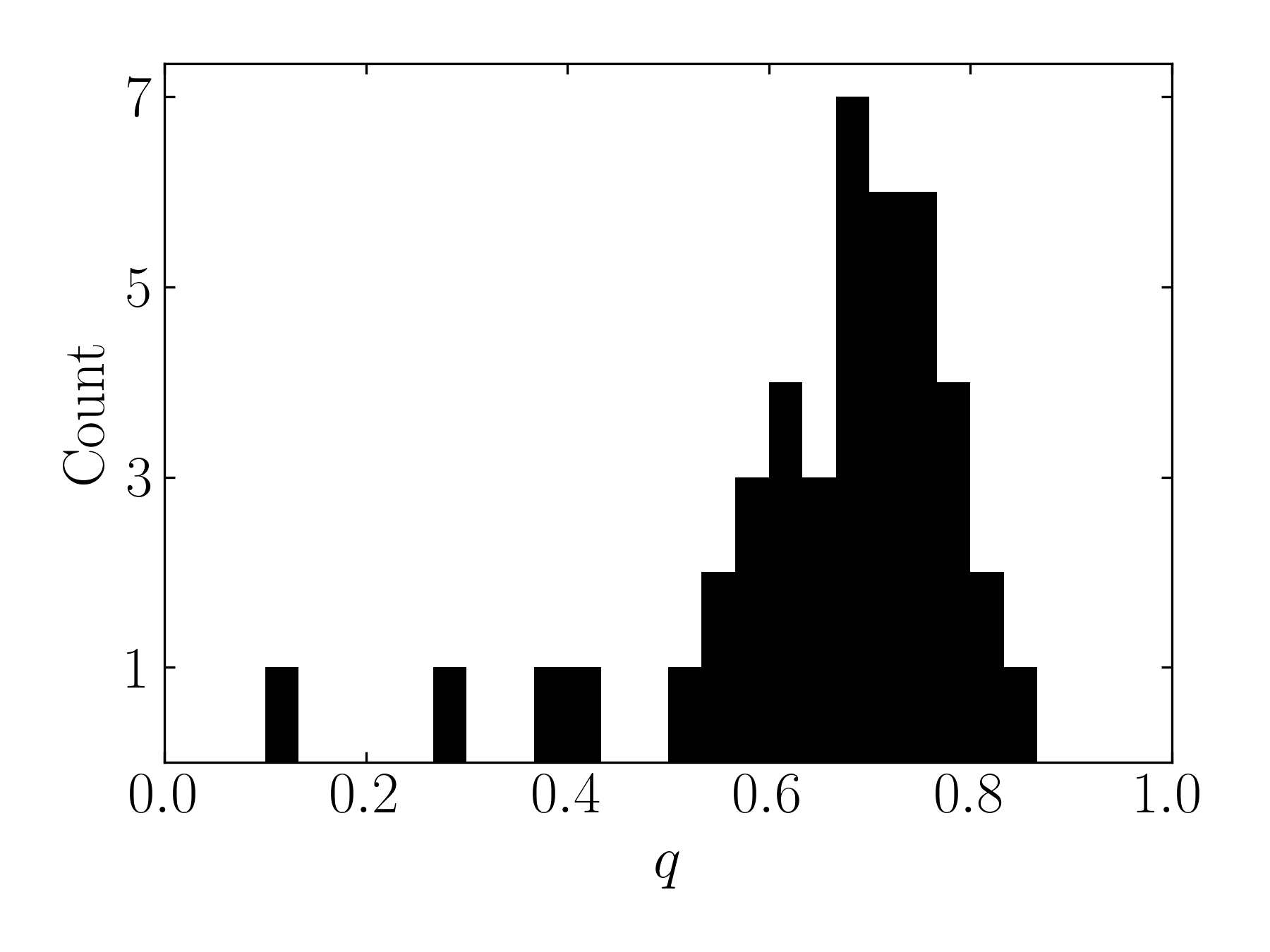}
    \caption{Histogram of the mass ratios $q \equiv m_2 / m_1$ of binary BH
    mergers in the O3a data release, excluding the two NS-NS mergers but including
    GW190814, whose $2.5M_{\odot}$ secondary may be a BH \citep{LIGOO3a}.
    }\label{fig:qhist}
\end{figure}

The fourth possible indicator of BH binary formation mechanisms is the
distribution of masses and mass ratios of merging BHs. In Fig.~\ref{fig:qhist},
we show the distribution of the mass ratio $q \equiv m_2 / m_1$, where $m_1 \geq
m_2$, for all LIGO/VIRGO binaries detected as of the O3a data release
\citep{LIGOO3a}\footnote{Note that Fig.~\ref{fig:qhist} should
not be interpreted as directly reflecting the distribution of merging BH
binaries, as there are many selection effects and observational biases, e.g.\
systems with smaller $q$ are harder to detect for the same $M_{\rm chirp}$ or
$m_{12}$. For a detailed statistical analysis, see \citet{LIGOO3a}.}. The
distribution distinctly peaks around $q \sim 0.7$. BH binaries formed via
isolated binary evolution are generally expected to have $q \gtrsim 0.5$
\citep{belczynski2016first, olejak2020}. On the other hand, dynamical formation
channels may produce a larger variety of distributions for the binary mass ratio
\citep[e.g.,][]{rodriguez2016binary, silsbee2017lidov, fragione2019}.

In this paper, we study in detail the mass ratio distribution for BH mergers
induced by tertiary companions in isolated triple systems. In this scenario, a tertiary
BH on a sufficiently inclined (outer) orbit induces phases of extreme
eccentricity in the inner binary via the von Zeipel-Lidov-Kozai
(ZLK;\@\citealp{zeipel, lidov, kozai}) effect, leading to efficient gravitational
radiation and orbital decay.  While the original ZLK effect relies on the
leading-order, quadrupolar gravitational perturbation from the tertiary on the
inner binary, the octupole order terms can become important \citep[sometimes
known as the eccentric Kozai mechanism, e.g.][]{naoz2016eccentric} when the
triple system is mildly hierarchical, the outer orbit is eccentric ($e_{\rm out}
\neq 0$) and the inner binary BHs have unequal masses
\citep[e.g.,][]{ford2000secular, blaes2002kozai, lithwick2011eccentric, LML15}.
The strength of the octupole effect depends on the dimensionless parameter
\begin{align}
    \epsilon_{\rm oct} = \frac{m_1 - m_2}{m_1 + m_2} \frac{a}{a_{\rm out}}
        \frac{e_{\rm out}}{1 - e_{\rm out}^2}.\label{eq:eps_oct}
\end{align}
where $a, a_{\rm out}$ are the semi-major axes of the inner and outer binaries,
respectively. Previous studies have shown that the octupole terms generally
increase the inclination window for extreme eccentricity excitation, and thus
enhance the rate of successful binary mergers \citep{LL18}. As $\epsilon_{\rm
oct} \propto (1-q)/(1+q)$ increases with decreasing $q$, we expect that
ZLK-induced BH mergers favor binaries with smaller mass ratios. The main goal of
this paper is to quantify the dependence of the merger fraction/probability on
$q$, using a combination of analytical and numerical calculations. We focus on
the cases where the tertiary mass is comparable to the binary BH masses. When
the tertiary mass $m_3$ is much larger than $m_{12}=m_1+m_2$ (as in the case of
a supermassive BH tertiary), dynamical stability of the triple requires $a_{\rm
out} (1-e_{\rm out}) /[a(1+e)]  \gtrsim 3.7 (m_3/m_{12})^{1/3} \gg 1$
\citep{kiseleva}, which implies that the octupole effect is negligible.

This paper is organized as follows. In Section~\ref{s:background}, we review
some analytical results of ZLK oscillations and examine how the octupole terms
affect the inclination window and probability for extreme eccentricity
excitation. In Section~\ref{s:with_gw}, we study tertiary-induced BH mergers
using a combination of numerical and analytical approaches. We propose new
semi-analytical criteria (Section~\ref{ss:nogw_merger}) that
allow us to determine, without full numerical integration, whether an initial BH
binary can undergo a ``one-shot merger'' or a more gradual merger induced by the
octupole effect of an tertiary. In Section~\ref{s:merger_frac}, we calculate the
merger fraction as a function of mass ratio for some representative triple
systems. In Section~\ref{s:q_dist}, we study the mass ratio distribution of the
initial BH binaries based on the properties of main-sequence (MS) stellar
binaries and the MS mass to BH mass mapping. Using the result of
Section~\ref{s:merger_frac}, we illustrate how the final merging BH binary mass
distribution may be influenced by the octupole effect for tertiary-induced
mergers. We summarize our results and their implications in
Section~\ref{s:conclusion}.

\section{Von Zeipel-Lidov-Kozai (ZLK) Oscillations: Analytical
Results}\label{s:background}

Consider two BHs orbiting each other with masses $m_1$ and $m_2$ on a orbit with
semi-major axis $a$, eccentricity $e$, and angular momentum $\bm{L}$. An external,
tertiary BH of mass $m_3$ orbits this inner binary with semi-major axis $a_{\rm
out}$, eccentricity $e_{\rm out}$, and angular momentum $\bm{L}_{\rm out}$. The
reduced masses of the inner and outer binaries are $\mu \equiv m_1m_2 / m_{12}$
and $\mu_{\rm out} \equiv m_{12} m_3 / m_{123}$ respectively, where $m_{12} =
m_1 + m_2$ and $m_{123} = m_{12} + m_3$. These two binary orbits are further
described by three angles: the inclinations $i$ and $i_{\rm out}$, the arguments
of pericenters $\omega$ and $\omega_{\rm out}$, and the longitudes of the
ascending nodes $\Omega$ and $\Omega_{\rm out}$. These angles are defined in a
coordinate system where the $z$ axis is aligned with the total angular momentum
$\bm{J} = \bm{L} + \bm{L}_{\rm out}$ (i.e., the invariant plane is perpendicular
to $\bm{J}$). The mutual inclination between the two orbits is denoted $I \equiv
i + i_{\rm out}$. Note that $\Omega_{\rm out} = \Omega + 180^\circ$.

To study the evolution of the inner binary under the influence of the tertiary
BH, we use the double-averaged secular equations of motion, including the
interactions between the inner binary and the tertiary up to the octupole level
of approximation as given by \citet{LML15}. Throughout this paper, we restrict
to hierarchical triple systems where the double-averaged secular equations are
valid -- systems 
with relatively small $a_{\rm out}/a$ may require solving the
single-averaged equations of motion or direct N-body integration
\citep[see][]{Antonini_2012, Antonini_2014, LuoKatzDong, LeiCirciOrtore,
bin_misc5, LL19, Hamers}\footnote{ Although we do not study such systems in this
paper, we expect that a qualitatively similar dependence of the merger
probability on the mass ratio remains, since the strength of the octupole effect
in the single-averaged secular equations is also proportional to $(1-q)/(1+q)$
\citep[see Eq.~25 of][]{bin_misc5}.}. For the remainder of this section, we
include general relativistic apsidal precession of the inner binary, a first
order post-Newtonian (1PN) effect, but omit the emission of GWs, a 2.5PN
effect -- this 
will be considered in Section~\ref{s:with_gw}. We group the results by
increasing order of approximation, starting by ignoring the octupole-order
effects entirely.

\subsection{Quadrupole Order}

At the quadrupole order, the tertiary induces eccentricity oscillations in the
inner binary on the characteristic timescale
\begin{align}
    t_{\rm ZLK} = \frac{1}{n}\frac{m_{12}}{m_3}
            \p{\frac{a_{\rm out, eff}}{a}}^3,\label{eq:def_tzlk}
\end{align}
where $n \equiv \sqrt{Gm_{12} / a^3}$ is the mean motion of the inner binary,
and $a_{\rm out, eff} \equiv a_{\rm out}\sqrt{1 - e_{\rm out}^2}$. During these
oscillations, there are two conserved quantities, the total energy and the total
orbital angular momentum. Through some manipulation, the total angular momentum
can be written in terms of the conserved quantity $K$ given by
\begin{align}
    K \equiv j(e) \cos I - \frac{\eta}{2}e^2.\label{eq:def_K}
\end{align}
Here, $j(e) \equiv \sqrt{1 - e^2}$ and $\eta$ is the ratio of the magnitudes of
the angular momenta at zero inner binary eccentricity:
\begin{align}
    \eta \equiv \p{\frac{L}{L_{\rm out}}}_{e = 0}
        = \frac{\mu}{\mu_{\rm out}}\s{\frac{m_{12}a}
            {m_{123}a_{\rm out}(1 - e_{\rm out}^2)}}^{1/2}.\label{eq:def_eta}
\end{align}
Note that when $\eta = 0$, $K$ reduces to the classical ``Kozai constant'', $K =
j(e) \cos I$.

The maximum eccentricity $e_{\max}$ attained in these ZLK oscillations can be
computed analytically at the quadrupolar order. It depends on the
``competition'' between the 1PN apsidal precession rate $\dot{\omega}_{\rm GR}$
and the ZLK rate $t_{\rm ZLK}^{-1}$. The relevant dimensionless parameter is
\begin{align}
    \epsilon_{\rm GR} \equiv \p{\dot{\omega}_{\rm GR} t_{\rm ZLK}}_{e = 0}
        = \frac{3Gm_{12}}{c^2} \frac{m_{12}}{m_3}\frac{a_{\rm out, eff}^3}{a^4}.
        \label{eq:def_epsgr}
\end{align}
It can then be shown that, for an initially circular inner binary, $e_{\max}$ is
related to the initial mutual inclination $I_{\rm 0}$ by \citep{LML15,
anderson2016formation}:
\begin{align}
    \frac{3}{8}\frac{j^2(e_{\max}) - 1}{j^2(e_{\max})}\Big[&
        5\p{\cos I_0 + \frac{\eta}{2}}^2
        - \p{3 + 4\eta \cos I_0 + \frac{9}{4}\eta^2}j^2(e_{\max})
            \nonumber\\
        &+ \eta^2 j^4(e_{\max})
    \Big] + \epsilon_{\rm GR}\s{1 - \frac{1}{j(e_{\max})}} = 0.
    \label{eq:emax_quad}
\end{align}
In the limit $\eta \to 0$ and $\epsilon_{\rm GR} \to 0$, we recover the
well-known result $e_{\max} = \sqrt{1 - (5/3) \cos^2 I_0}$. For general $\eta$,
$e_{\max}$ attains its limiting value $e_{\lim}$ when $I_{\rm 0} = I_{\rm 0,
\lim}$, where \citep[see also][]{hamers_tp}
\begin{align}
    \cos I_{\rm 0, \lim} = \frac{\eta}{2}\s{\frac{4}{5}j^2(e_{\lim}) -
        1}.\label{eq:def_I0lim}
\end{align}
Note that $I_{\rm 0, \lim} \geq 90^\circ$ with equality only when $\eta
= 0$. Substituting Eq.~\eqref{eq:def_I0lim} into Eq.~\eqref{eq:emax_quad}, we find
that $e_{\lim}$ satisfies
\begin{align}
    \frac{3}{8}\s{j^2(e_{\lim}) - 1}&\s{-3 + \frac{\eta^2}{4}
        \p{\frac{4}{5}j^2(e_{\lim}) - 1}}\nonumber\\
        &+ \epsilon_{\rm GR}\s{1 - \frac{1}{j(e_{\lim})}} = 0.
        \label{eq:def_elim}
\end{align}
On the other hand, eccentricity excitation ($e_{\max} \geq 0$) is only possible
when $(\cos I_{\rm 0})_- \leq \cos I_{\rm 0} \leq (\cos I_{\rm 0})_+$ where
\begin{align}
    \p{\cos I_{\rm 0}}_{\pm} = \frac{1}{10}\p{-\eta \pm \sqrt{\eta^2 + 60 -
        \frac{80}{3}\epsilon_{\rm GR}}}.\label{eq:I0bounds}
\end{align}
For $I_{\rm 0}$ outside of this range, no eccentricity excitation is possible.
This condition reduces to the well-known $\cos^2 I_{\rm 0} \leq 3/5$ when $\eta
= \epsilon_{\rm GR} = 0$.

\subsection{Octupole Order: Test-particle Limit}\label{ss:oct_tp}

\begin{figure}
    \centering
    \includegraphics[width=\columnwidth]{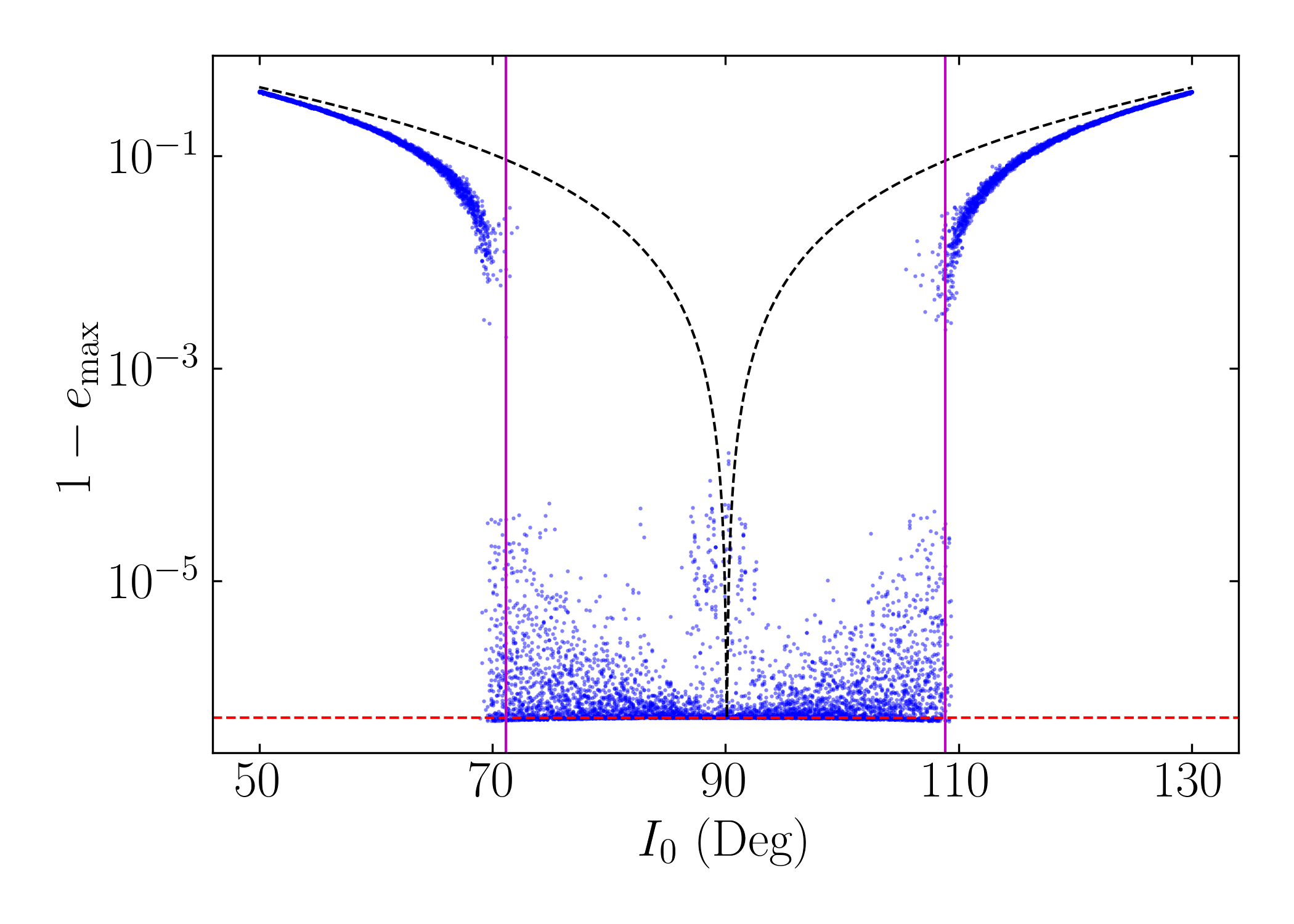}
    \caption{The maximum eccentricity achieved for an inner binary in the
    test-particle limit as a function of the initial inclination angle $I_0$.
    The triple system parameters are: $a = 100\;\mathrm{AU}$, $a_{\rm out, eff}
    = 3600\;\mathrm{AU}$, $m_{12} = 50M_{\odot}$, $m_3 = 30M_{\odot}$, and
    $e_{\rm out} = 0.6$; the corresponding octupole strength parameter is
    $\epsilon_{\rm oct} = 0.02$ and $\eta \simeq 0$. The octupole-level secular
    equations of motion are integrated for $2000t_{\rm ZLK}$ (see
    Eq.~\ref{eq:def_tzlk}), and the maximum eccentricity attained during this
    time is recorded and shown as a blue dot for each initial condition. We
    consider $1000$ initial inclinations in the range $50^\circ \leq I_0 \leq
    130^\circ$, and each $I_0$ is simulated five times, with the initial orbital
    elements $\omega$, $\omega_{\rm out}$, and $\Omega = \Omega_{\rm out} - \pi$
    chosen randomly $\in [0, 2\pi)$ 
    for each simulation. The dotted black line shows the quadrupole-level result
    (Eq.~\ref{eq:emax_quad} with $\eta = 0$), and $e_{\lim}$
    (Eq.~\ref{eq:def_elim}) is shown as the horizontal red line. The vertical
    purple lines denote the boundary of the octupole-active inclination window,
    based on the fitting formula from \citet{MLL16} (Eq.~\ref{eq:I_oct_MLL}).
    }\label{fig:composite_tp}
\end{figure}

The relative strength of the octupole-order potential to the quadrupole-order
potential is determined by the dimensionless parameter $\epsilon_{\rm oct}$
(Eq.~\ref{eq:eps_oct}). When $\epsilon_{\rm oct}$ is non-negligible, $K$ is no
longer conserved, and the system evolution becomes chaotic
\citep{ford2000secular, katz2011long, lithwick2011eccentric, li2014chaos,
LML15}. As a result, analytical (and semi-analytical) results have only been
given for the test-particle limit, where $m_2 = \eta = 0$. We briefly review
these results below.

Due to the non-conservation of $K$, $e_{\max}$ evolves irregularly ZLK cycles,
and the orbit may even flip between prograde ($I < 90^\circ$) and retrograde ($I
> 90^\circ$) if $K$ changes sign (in the test-particle limit, $K = j(e) \cos
I$). During these orbit flips, the eccentricity maxima reach their largest
values but do not exceed $e_{\lim}$ \citep{lithwick2011eccentric, LML15,
anderson2016formation}. These orbit flips occur on characteristic timescale
$t_{\rm ZLK, oct}$, given by \citep{antognini2015timescales}
\begin{align}
    t_{\rm ZLK, oct} = t_{\rm ZLK}\frac{128\sqrt{10}}{
        15\pi\sqrt{\epsilon_{\rm oct}}}.\label{eq:def_tzlkoct}
\end{align}
The octupole potential tends to widen the inclination range for which the
eccentricity can reach $e_{\lim}$; we refer to this widened range as the
\emph{octupole-active window}. Figure~\ref{fig:composite_tp} shows the maximum
eccentricity attained by an inner binary orbited by a tertiary companion with
inclination $I_0$. The octupole-active window is visible as a range of
inclinations centered on $I_0 = 90^\circ$ that attain $e_{\lim}$ (the red
horizontal dashed line in Fig.~\ref{fig:composite_tp}). \citet{katz2011long}
show that this window can be approximated using analytical arguments when
$\epsilon_{\rm oct} \ll 1$. \citet{MLL16} give a more general numerical fitting
formula describing the octupole-active window for arbitrary $\epsilon_{\rm
oct}$. They find that orbit flips and extreme eccentricity excitation occur for
$I_{\rm flip, -} \lesssim I_0 \lesssim I_{\rm flip, +}$ where
\begin{align}
    \cos^2 I_{\rm flip, \pm} = \begin{cases}
        0.26\p{\frac{\epsilon_{\rm oct}}{0.1}}
            - 0.536\p{\frac{\epsilon_{\rm oct}}{0.1}}^2\\
            \quad + 12.05\p{\frac{\epsilon_{\rm oct}}{0.1}}^3
            - 16.78\p{\frac{\epsilon_{\rm oct}}{0.1}}^4
            & \epsilon_{\rm oct} \lesssim 0.05,\\
        0.45 & \epsilon_{\rm oct} \gtrsim 0.05.
    \end{cases} \label{eq:I_oct_MLL}
\end{align}
In Fig.~\ref{fig:composite_tp}, we see that with the octupole effect included,
$e_{\max}$ indeed attains $e_{\lim}$ when $I_0$ is within the broad
octupole-active window given by Eq.~\eqref{eq:I_oct_MLL} (denoted by the
vertical purple lines in Fig.~\ref{fig:composite_tp}).

\subsection{Octupole Order: General Masses}\label{ss:oct_gen}

\begin{figure}
    \centering
    \includegraphics[width=\columnwidth]{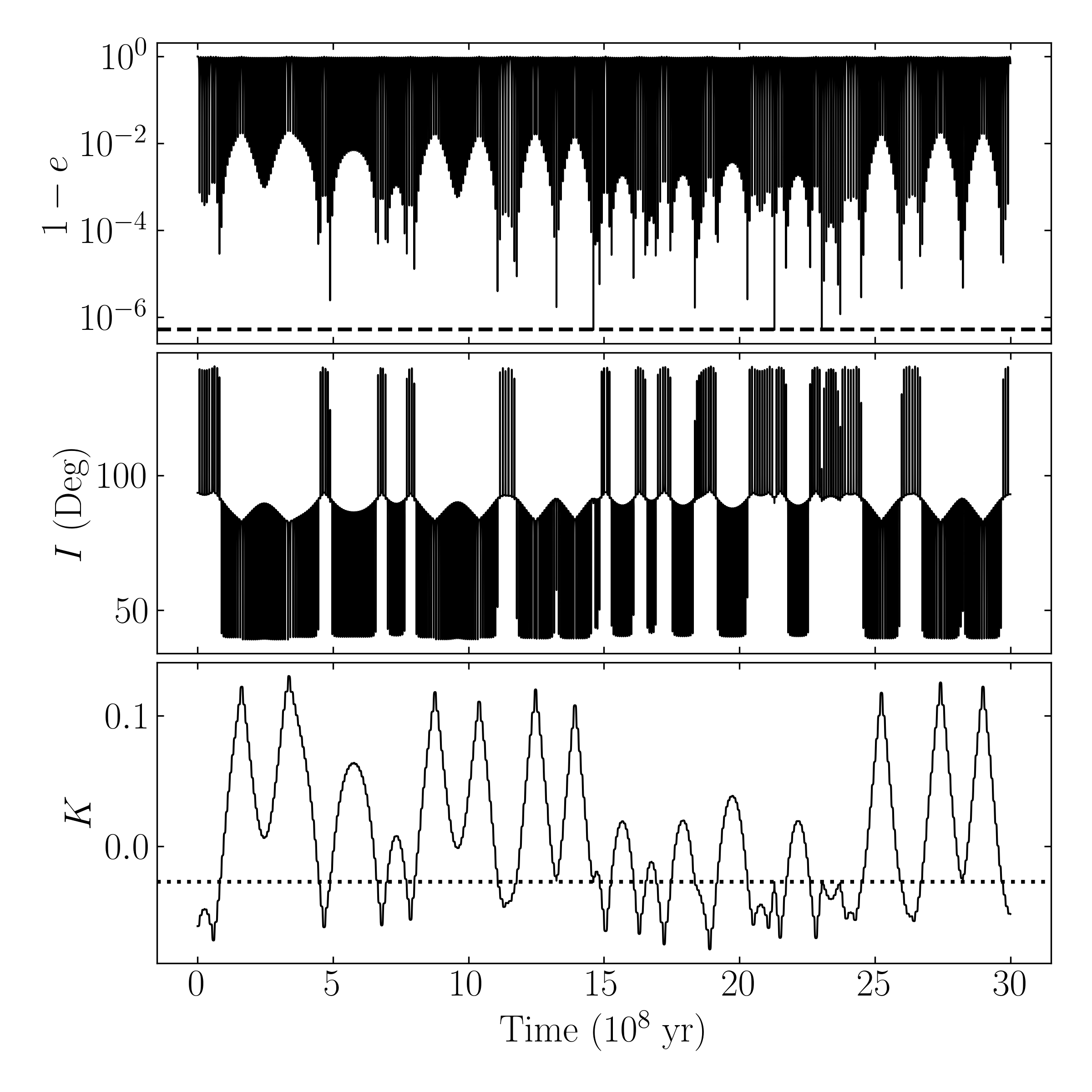}
    \caption{An example of the triple evolution for a system with significant
    octupole effects and finite $\eta$ (see Eq.~\ref{eq:def_eta}). We use the
    same system parameters as in Fig.~\ref{fig:composite_tp} except for $q =
    0.2$, corresponding to $\eta \approx 0.087$ and $\epsilon_{\rm oct} \approx
    0.007$, and $I_0 = 93.5^\circ$. The three panels show the inner orbit
    eccentricity, the mutual inclination, and the generalized ``Kozai constant''
    $K$ (Eq.~\ref{eq:def_K}). In the first panel, $e_{\lim}$ is denoted by the
    black dashed line. By comparing the second and third panels, we see that
    orbit flips occur when $K$ crosses the dotted line, given by $K = K_{\rm c}
    \equiv -\eta / 2$. }\label{fig:nogw_fiducial}
\end{figure}

For general inner binary masses, when the angular momentum ratio $\eta$ is
non-negligible, the octupole-level ZLK behavior is less well-studied
\citep[see][]{LML15}. Figure~\ref{fig:nogw_fiducial} shows an example of the
evolution of a triple system with significant $\eta$ and $\epsilon_{\rm oct}$.
Many aspects of the evolution discussed in Section~\ref{ss:oct_tp} are still
observed: the ZLK eccentricity maxima and $K$ evolve over timescales $\gg t_{\rm
ZLK}$; the eccentricity never exceeds $e_{\lim}$; when $K$ crosses $K_{\rm c}
\equiv -\eta / 2$, an orbit flip occurs (this follows by inspection of
Eq.~\ref{eq:def_K}).

\begin{figure}
    \centering
    \includegraphics[width=\columnwidth]{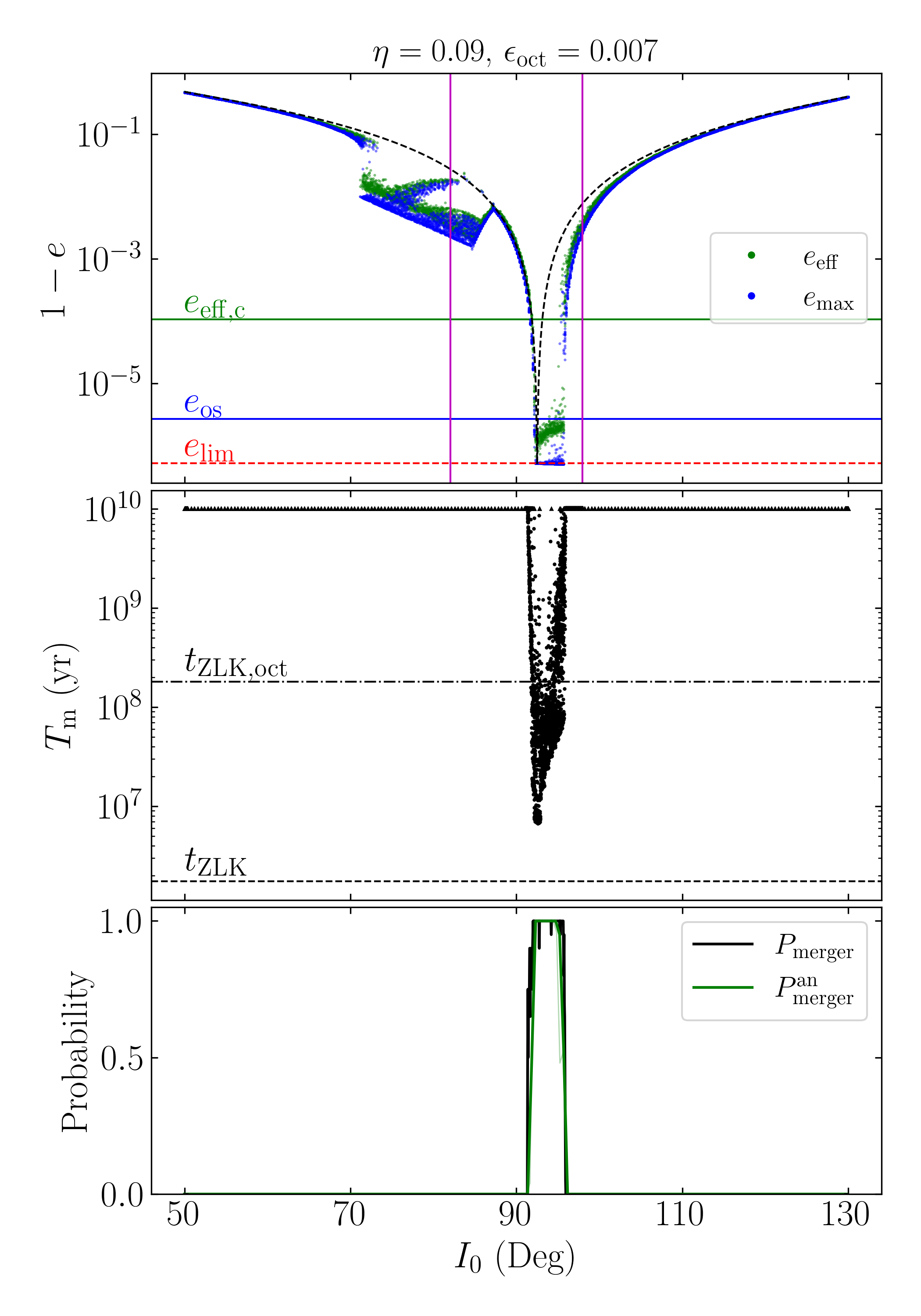}
    \caption{Eccentricity excitation and merger windows for the fiducial BH
    triple system ($a = 100\;\mathrm{AU}$, $a_{\rm out, eff} =
    3600\;\mathrm{AU}$, $m_{12} = 50M_{\odot}$, $m_3 = 30M_{\odot}$) with $q =
    0.5$ and $e_{\rm out} = 0.6$, corresponding to $\eta \approx 0.087$ and
    $\epsilon_{\rm oct} \approx 0.007$. In the top panel, for each of $1000$
    initial inclinations, we choose $5$ different random $\omega$, $\omega_{\rm
    out}$, and $\Omega$ as initial conditions and evolve the system for
    $2000t_{\rm ZLK}$ without GW radiation. The effective eccentricity $e_{\rm
    eff}$ (Eq.~\ref{eq:def_e_eff}; green dots) as well as the maximum
    eccentricity $e_{\max}$ (blue dots) over this period are displayed. For
    comparison, $e_{\rm eff, c}$ (Eq.~\ref{eq:def_e_eff_c}) is given by the
    horizontal green dashed line, $e_{\rm os}$ (Eq.~\ref{eq:def_e_os}) is shown
    as the horizontal blue line, and $e_{\lim}$ (Eq.~\ref{eq:def_elim}) is shown
    as the horizontal red dashed line. The vertical purple lines denote the
    test-mass octupole-active window and are given by the fitting formula of
    \citet{MLL16}; they do not longer accurately describe the
    $e_{\lim}$-attaining inclination window because $\eta$ is finite. The black
    dashed line is is the quadrupole-level result as given by
    Eq.~\eqref{eq:emax_quad}. In the middle panel, we show the binary merger
    times when including GW radiation and using the same range of initial
    conditions. Numerical integrations are terminated when $T_{\rm m} >
    10\;\mathrm{Gyr}$ and marked as unsuccessful mergers. The horizontal dashed
    line denotes $t_{\rm ZLK}$ (Eq.~\ref{eq:def_tzlk}) while the horizontal
    dash-dotted line indicates $t_{\rm ZLK, oct}$ (Eq.~\ref{eq:def_tzlkoct}).
    Here, each $I_{\rm 0}$ is run $20$ times with uniform distributions of
    $\omega$, $\omega_{\rm out}$, and $\Omega$, so we can estimate the merger
    probability $P_{\rm merger}$ (Eq.~\ref{eq:def_pmerge}) for each $I_{\rm
    0}$ -- $P_{\rm merger}$ 
    is shown as the black line in the bottom panel. As described in
    Section~\ref{ss:nogw_merger}, the merger probability can be predicted
    semi-analytically using the results of the top panel and
    Eq.~\eqref{eq:def_pmerge_sa}, and is denoted by $P_{\rm merger}^{\rm an}$.
    In the bottom panel, the thick green line shows $P_{\rm merger}^{\rm an}$
    when using an integration time of $2000 t_{\rm ZLK} \approx 3\;\mathrm{Gyr}$
    for the non-dissipative simulations, and thin green line shows the
    prediction using an integration time of $500 t_{\rm ZLK}$. The agreement of
    $P_{\rm merger}^{\rm an}$ with $P_{\rm merger}$ is good and improves when
    using the longer integration time. }\label{fig:composite_dist}
\end{figure}

\begin{figure}
    \centering
    \includegraphics[width=\columnwidth]{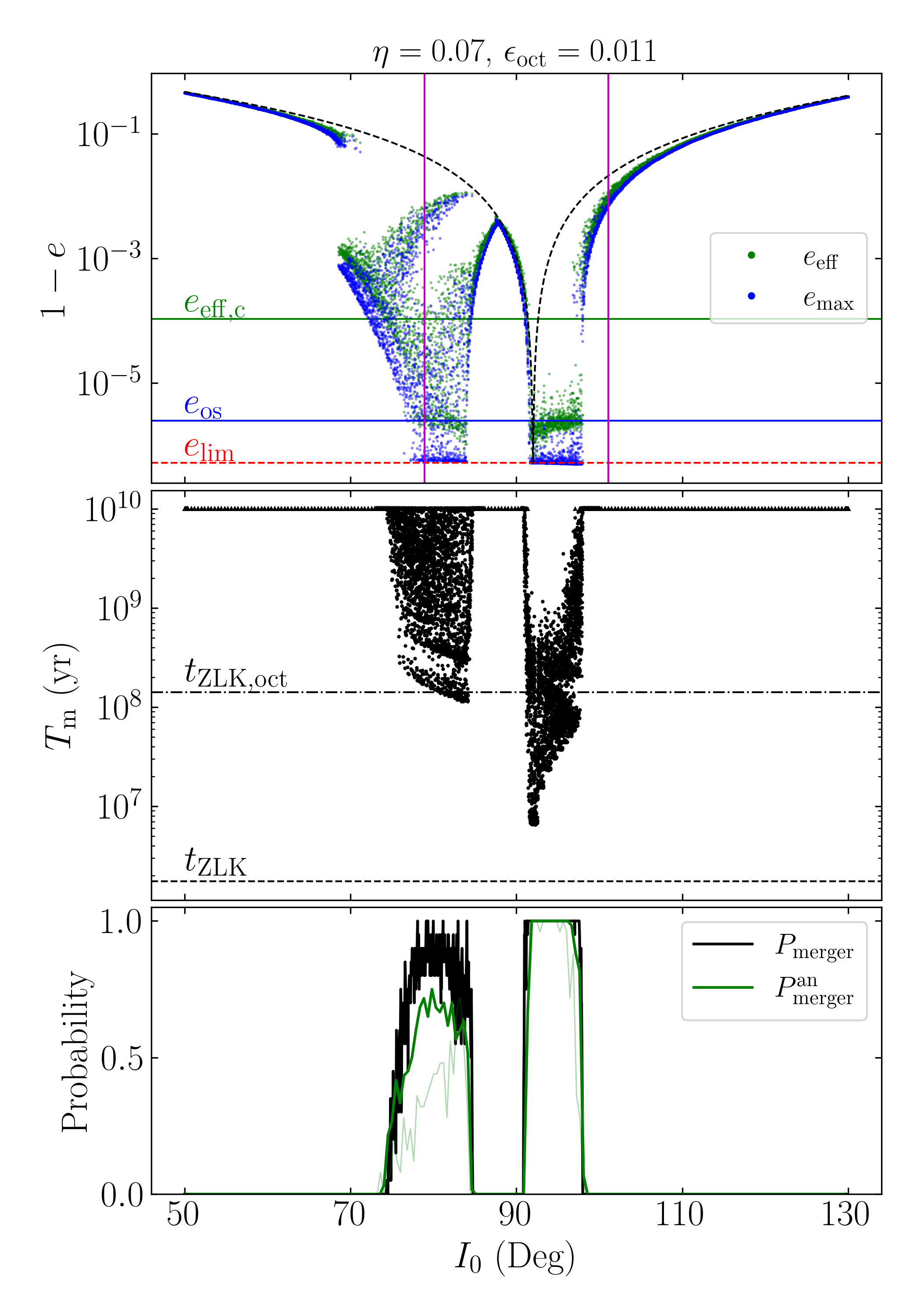}
    \caption{Same as Fig.~\ref{fig:composite_dist} but for $q = 0.3$,
    corresponding to $\eta \approx 0.07$ and $\epsilon_{\rm oct} \approx
    0.011$.}\label{fig:composite_1p3}
\end{figure}

\begin{figure}
    \centering
    \includegraphics[width=\columnwidth]{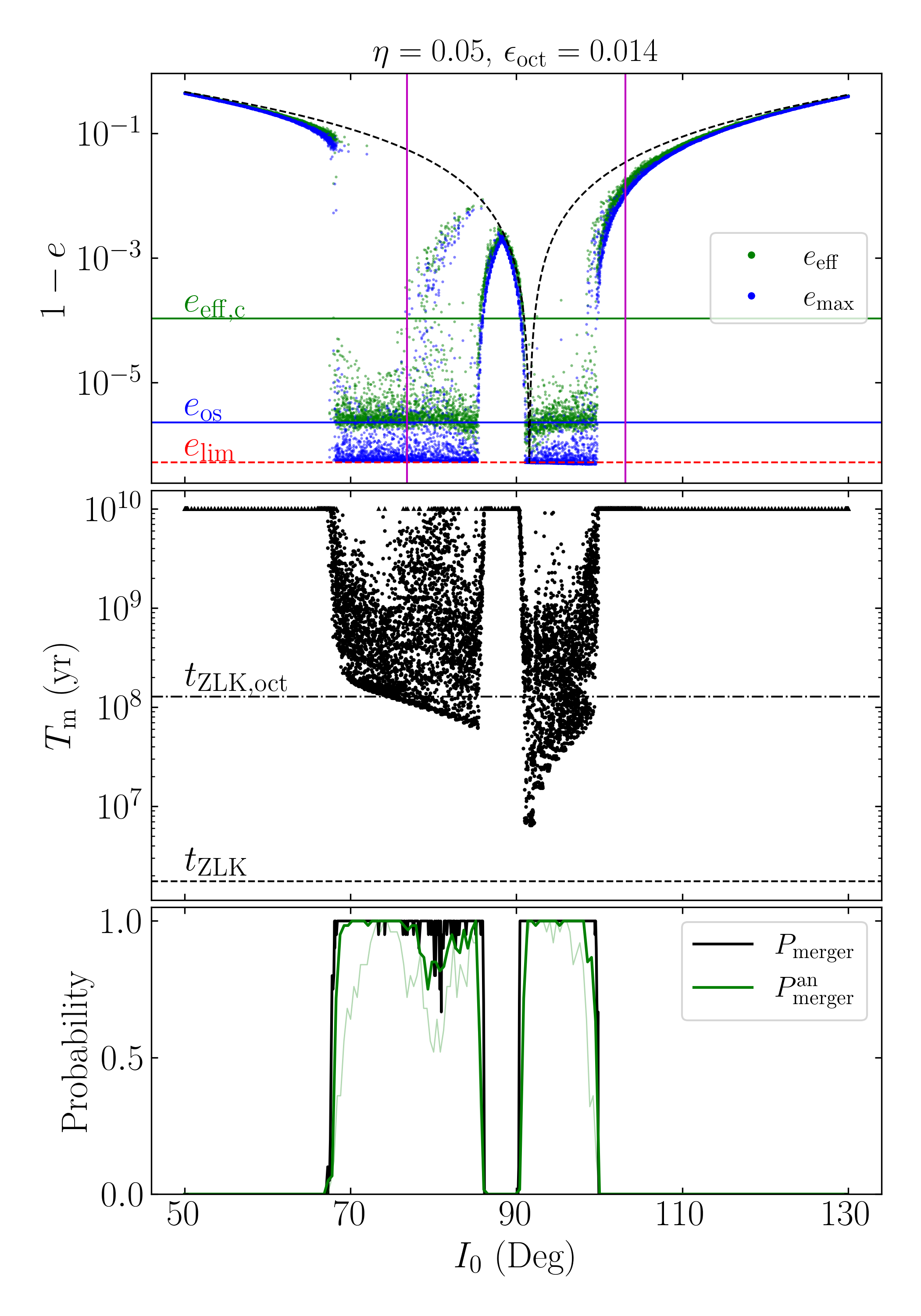}
    \caption{Same as Fig.~\ref{fig:composite_dist} but for $q = 0.2$,
    corresponding to $\eta \approx 0.054$ and $\epsilon_{\rm oct} \approx
    0.014$. }\label{fig:composite_1p2}
\end{figure}

\begin{figure}
    \centering
    \includegraphics[width=\columnwidth]{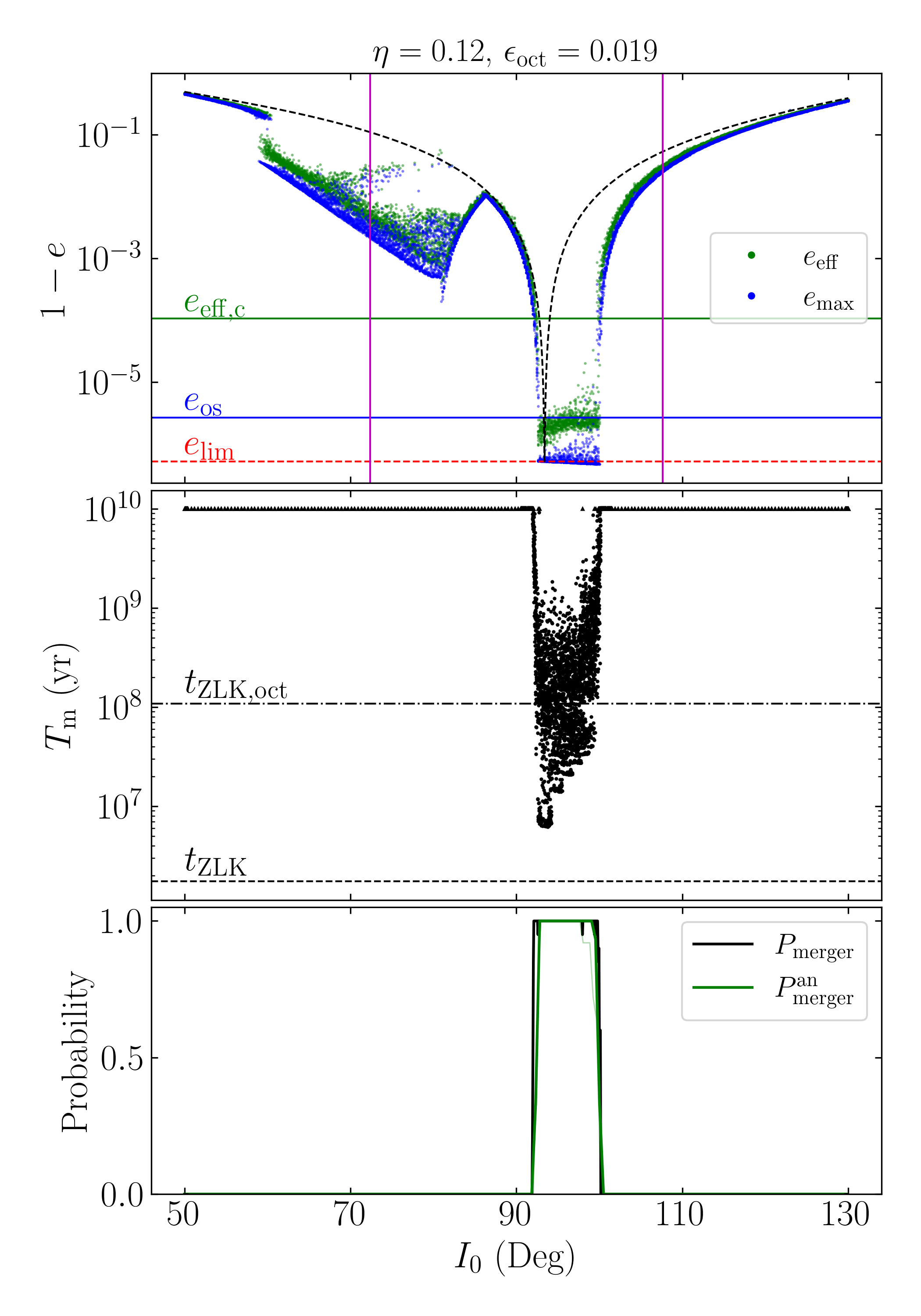}
    \caption{Same as Fig.~\ref{fig:composite_dist} but for $e_{\rm out} = 0.9$
    while holding $a_{\rm out, eff}$ the same, corresponding to $\eta =
    0.118$ and $\epsilon_{\rm oct} = 0.019$. }\label{fig:composite_e91p5}
\end{figure}

\begin{figure}
    \centering
    \includegraphics[width=\columnwidth]{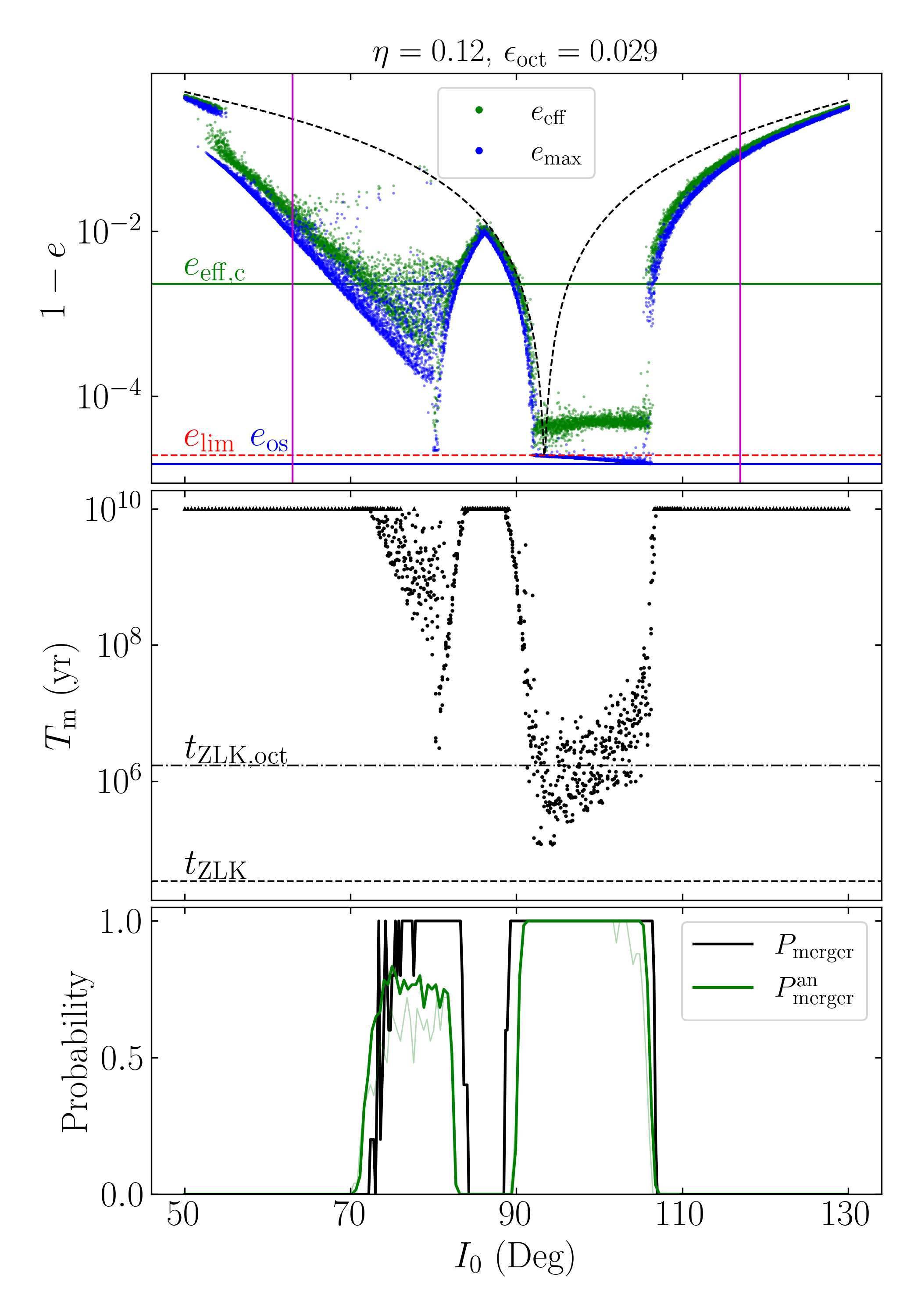}
    \caption{Same as Fig.~\ref{fig:composite_dist} but for a more compact inner
    binary; the parameters are $a_0 = 10\;\mathrm{AU}$, $a_{\rm out, eff} =
    700\;\mathrm{AU}$, $m_{12} = 50M_{\odot}$, $m_3 = 30M_{\odot}$, $e_{\rm
    out} = 0.9$, and $q = 0.4$, corresponding to $\eta = 0.118$ and
    $\epsilon_{\rm oct} = 0.029$. Here, $P_{\rm merger}$ is computed with
    only $5$ integrations (for random $\omega$, $\omega_{\rm out}$, and
    $\Omega$) for each $I_0$.}\label{fig:composite_bindist}
\end{figure}

However, Eq.~\eqref{eq:I_oct_MLL} no longer describes the octupole-active window
as $\eta$ is non-negligible \citep[see also][]{rodet_inprep}. In the top panel
of Fig.~\ref{fig:composite_dist}, the blue dots show the maximum achieved
eccentricity of a system with the same parameters as Fig.~\ref{fig:composite_tp}
except with $q = 0.5$ (so $\epsilon_{\rm oct} = 0.007$ and $\eta = 0.087$).
Here, it can be seen that no prograde systems can attain $e_{\lim}$, and only a
small range of retrograde inclinations $\geq I_{\rm 0, \lim}$ (see
Eq.~\ref{eq:def_I0lim}) are able to reach $e_{\lim}$. In fact, there is even a
clear double valued feature around $I \approx 75^\circ$ in the top panel of
Fig.~\ref{fig:composite_dist} that is not present in
Fig.~\ref{fig:composite_tp}. If $q$ is decreased to $0.3$
(Fig.~\ref{fig:composite_1p3}) or further to $0.2$
(Fig.~\ref{fig:composite_1p2}), $\epsilon_{\rm oct}$ increases while $\eta$
decreases. This permits a larger number of prograde systems to reach $e_{\lim}$,
though a small range of inclinations near $I_0 = 90^\circ$ still do not reach
$e_{\lim}$; we call this range of inclinations the ``octupole-inactive gap''. On
the other hand, if $q$ is held at $0.5$ as in Fig.~\ref{fig:composite_dist} and
$e_{\rm out}$ is increased to $0.9$ while holding $a_{\rm out, eff} =
3600\;\mathrm{AU}$ constant, both $\epsilon_{\rm oct}$ and $\eta$ increase; the
top panel of Fig.~\ref{fig:composite_e91p5} shows that prograde systems still
fail to reach $e_{\lim}$ for these parameters, despite the increase in
$\epsilon_{\rm oct}$. The top panel of Fig.~\ref{fig:composite_bindist}
illustrates the behavior when the inner binary is substantially more compact ($a
= 10\;\mathrm{AU}$): even though $\epsilon_{\rm oct}$ is larger than it is
in any of Figs.~\ref{fig:composite_dist}--\ref{fig:composite_e91p5}, we
see that prograde perturbers fail to attain $e_{\lim}$. All of these examples
(top panels of Figs.~\ref{fig:composite_dist}--\ref{fig:composite_bindist})
illustrate importance of $\eta$ in determining the range of inclinations for the
system to be able to reach $e_{\lim}$.

In general, we find that a symmetric octupole-active window (as in
Eq.~\ref{eq:I_oct_MLL}) can be realized for sufficiently small $\eta$.
\citet{rodet_inprep} considered some examples of triple systems (consisting of
MS stars with planetary companions and tertiaries, for which the short-range
forces is dominated by tidal interaction) and found that $\eta\lesssim 0.1$ is
sufficient for a symmetric octupole-active window. In the cases considered in
this paper, a smaller $\eta$ is necessary (e.g., $\eta\simeq 0.054$ in
Fig.~\ref{fig:composite_1p2}). Thus, the critical $\eta$ above which the
symmetry of the octupole-active window is significantly broken likely depends on
the dominant short-range forces and $e_{\lim}$ [in \citet{rodet_inprep}, $1 -
e_{\lim} \sim 10^{-3}$, while in Figs.~\ref{fig:composite_tp}
and~\ref{fig:composite_dist}--\ref{fig:composite_bindist}, $1 - e_{\lim}
\lesssim 10^{-5}$]. In general, when $\eta$ is non-negligible, there are up to
two octupole-active windows: a prograde window whose existence depends on the
specific values of $\eta$ and $\epsilon_{\rm oct}$, and a retrograde window that
always exists.


\section{Tertiary-Induced Black Hole Mergers}\label{s:with_gw}

Emission of gravitational waves (GWs) affects the evolution of the inner
binary, which can be incorporated into the secular equations of motion for the
triple \citep[e.g.,][]{peters1964, LL18}. The associated orbital and eccentricity
decay rates are \citep{peters1964}:
\begin{align}
    \at{\frac{1}{a}\rd{a}{t}}_{\rm GW} &\equiv -\frac{1}{t_{\rm GW}}\nonumber\\
        &= -\frac{64}{5}\frac{G^3 \mu m_{12}^2}{c^5a^4j^7(e)}
            \p{1 + \frac{73}{24}e^2 + \frac{37}{96}e^4}\label{eq:def_tgw},\\
    \at{\rd{e}{t}}_{\rm GW} &= -\frac{304}{15}\frac{G^3 \mu m_{12}^2}{c^5a^4}
        \frac{1}{j^{5}(e)}\p{1 + \frac{121}{304}e^2}\label{eq:dedt_gw}.
\end{align}
GW emission can cause the orbit to decay significantly when extreme
eccentricities are reached during the ZLK cycles described in the previous
section. This allows even wide binaries ($\sim 100\;\mathrm{AU}$) to merge
efficiently within a Hubble time. While various numerical examples of such
tertiary-induced mergers have been given before (e.g., \citet{LL18}; see also
\citet{LL19} for ``population synthesis''), in this section we examine the
dynamical process in detail in order to develop an analytical understanding. Our
fiducial system parameters are as in Fig.~\ref{fig:nogw_fiducial}: $a_{\rm out,
eff} = 4500\;\mathrm{AU}$, $e_{\rm out} = 0.6$, $m_{12} = 50M_{\odot}$ (with
varying $q$), $m_3 = 30M_{\odot}$, and the inner binary has initial $a_0 =
100\;\mathrm{AU}$ and $e_0 = 10^{-3}$.

\subsection{Merger Windows and Probability: Numerical Results}\label{ss:windows}

To understand what initial conditions lead to successful mergers within a Hubble
time, we integrate the double-averaged octupole-order ZLK equations including GW
radiation. We terminate each integration if either $a = 0.005a_0$ (a successful
merger) or the system age reaches $10\;\mathrm{Gyr}$. We can
verify that the inner binary is effectively decoupled from the tertiary for this
orbital separation by evaluating $\epsilon_{\rm GR}$ (Eq.~\ref{eq:def_epsgr}):
\begin{equation}
    \epsilon_{\rm GR} = 1.8 \times 10^6
        \p{\frac{m_{12}}{50 M_{\odot}}}^2
        \p{\frac{a_{\rm out, eff}}{3600\;\mathrm{AU}}}^3
        \p{\frac{m_3}{30M_{\odot}}}^{-1}
        \p{\frac{a}{0.5\;\mathrm{AU}}}^{-4}.
        \label{eq:epsgr_0005}
\end{equation}
The middle panel of Fig.~\ref{fig:composite_dist} shows the merger time $T_{\rm
m}$ as a function of $I_0$ for our fiducial parameters with $q = 0.5$. We note
that only retrograde inclinations lead to successful mergers, and almost all
successful mergers are rapid, with $T_{\rm m} \sim t_{\rm ZLK, oct}$. These are
the result of a system merging by emitting a single large burst of GW radiation
during an extreme-eccentricity ZLK cycle, which we term a ``\emph{one-shot
merger}''\footnote{It is important to note that these ``one-shot
mergers'' are distinct from the ``fast'' mergers previously discussed in the
literature \citep[e.g.][]{wen2003eccentricity, randall2018analytical,
su2020spin}: The one-shot mergers discussed here occur when the
maximum eccentricity attained by the inner binary over an \emph{octupole} cycle
(i.e.\ within the first $\sim t_{\rm ZLK, oct}$) is sufficiently large to
produce a prompt merger, while the references cited above neglect octupole-order
effects and study the scenario when the maximum eccentricity attained in a
\emph{quadrupole} ZLK cycle (i.e.\ within the first $\sim t_{\rm ZLK}$) is
sufficiently large to produce a prompt merger. When the octupole effect is
non-negligible, it can drive systems to much more extreme eccentricities than
can the quadrupole-order effects alone (compare the blue dots and black dashed
line in Fig.~\ref{fig:composite_dist}), and thus our ``one-shot mergers'' occur
for a larger range of $I_0$ than do quadrupole-order ``fast'' mergers.}. In
Fig.~\ref{fig:composite_1p3}, $q$ is decreased to $0.3$, and some prograde
systems are also able to merge successfully. However, these prograde systems
exhibit a broad range of merger times, with $T_{\rm m} \gtrsim t_{\rm ZLK,
oct}$. These occur when a system gradually emits a small amount of GW radiation
at every eccentricity maximum -- we term this a 
``\emph{smooth merger}''. Additionally, the octupole-inactive gap near $I_0 =
90^\circ$ is visible in the merger time plot (middle panel of
Fig.~\ref{fig:composite_1p3}). The middle panels of
Figs.~\ref{fig:composite_1p2}--\ref{fig:composite_bindist} show the behavior of
$T_{\rm m}$ for the other parameter regimes and also exhibit these two
categories of mergers and the octupole-inactive gap.

Due to the chaotic nature of the octupole-order ZLK effect, the initial
inclination $I_0$ alone is not sufficient to determine with certainty whether a
system can merge within a Hubble time. Instead, for a given $I_0$, we can use
numerical integrations with various $\omega$, $\omega_{\rm out}$, and $\Omega$
to compute a merger probability, denoted by
\begin{align}
    P_{\rm merger}\p{I_0; q, e_{\rm out}} = P\p{T_{\rm m} < 10\;\mathrm{Gyr}},
        \label{eq:def_pmerge}
\end{align}
where the notation $P_{\rm merger}\p{I_0; q, e_{\rm out}}$ highlights the
dependence of $P_{\rm merger}$ on $q$ and $e_{\rm out}$, two of the key factors
that determine the strength of the octupole effect (of course $P_{\rm merger}$
depends on other system parameters such as $m_{12}$, $a_0$, $a_{\rm out}$,
etc.). The bottom panels of
Figs.~\ref{fig:composite_dist}--\ref{fig:composite_bindist} show our numerical
results. In all of these plots, there is a retrograde inclination window for
which successful merger is guaranteed. In Fig.~\ref{fig:composite_1p3}, it can
be seen that a large range of prograde inclinations have a probabilistic
outcome. In Fig.~\ref{fig:composite_1p2}, while the enhanced octupole strength
allows for most of the prograde inclinations to merge with certainty, there is
still a region around $I_0 \approx 80^\circ$ where $P_{\rm merger} < 1$.

\subsection{Merger Probability: Semi-analytic
Criteria}\label{ss:nogw_merger}

By comparing the top and bottom panels of
Figs.~\ref{fig:composite_dist}--\ref{fig:composite_bindist}, it is clear that
their features are correlated: in all five cases, the retrograde merger window
occupies the same inclination range as the retrograde octupole-active window,
while $P_{\rm merger}$ is only nonzero for prograde inclinations where
$e_{\max}$ nearly attains $e_{\lim}$. Here, we further develop this connection
and show that the non-dissipative simulations can be used to predict the
outcomes of simulations with GW dissipation rather reliably.

In Section~\ref{ss:windows}, we identified both one-shot and smooth mergers in
our simulations. Towards understanding the one-shot mergers, we first define
$e_{\rm os}$ to be the $e_{\max}$ required to dissipate an order-unity fraction
of the binary's orbital energy via GW emission in a single ZLK cycle. Since a
binary spends a fraction $\sim j(e_{\max})$ of each ZLK cycle near $e_{\max}$
\citep[e.g.,][]{anderson2016formation}, we set
\begin{align}
    j\p{e_{\rm os}}\at{\rd{\ln a}{t}}_{e = e_{\rm os}} =
        -\frac{1}{t_{\rm ZLK}},
\end{align}
where $\rdil{(\ln a)}{t}$ is given by Eq.~\eqref{eq:def_tgw}. This yields
\begin{align}
    j^6(e_{\rm os}) \equiv \frac{425 t_{\rm ZLK}}{96t_{\rm GW, 0}}
        =
        \frac{170}{3}
            \frac{G^3 \mu m_{12}^3}{m_3c^5a^4n}
            \p{\frac{a_{\rm out, eff}}{a}}^3,
            \label{eq:def_e_os}
\end{align}
where $t_{\rm GW, 0} = (t_{\rm GW})_{e = 0}$ (see Eq.~\ref{eq:def_tgw}) is given
by
\begin{align}
    t_{\rm GW, 0}^{-1} = \frac{64}{5}\frac{G^3 \mu m_{12}^2}{c^5a^4},
\end{align}
we have approximated $e_{\rm os} \approx 1$. Eq.~\eqref{eq:def_e_os} is
equivalent to
\begin{align}
    1 - e_{\rm os} \approx{}& 3 \times 10^{-6}
        \p{\frac{m_{12}}{50M_{\odot}}}^{7/6}
        \p{\frac{q / (1 + q)^2}{1/4}}^{1/3}
        \p{\frac{m_3}{30M_{\odot}}}^{-1/3}\nonumber\\
        &\times \p{\frac{a_{\rm out, eff}}{3600 \;\mathrm{AU}}}
            \p{\frac{a}{100\;\mathrm{AU}}}^{-11/6}.
\end{align}
Then, if a system satisfies $e_{\max} > e_{\rm os}$ with $e_{\max}$ based on
non-dissipative integration, it is expected attain a sufficiently large
eccentricity to undergo a one-shot merger.

Towards understanding smooth mergers, we seek a characteristic eccentricity that
captures GW emission over many ZLK cycles. We define $e_{\rm eff}$ as an
effective ZLK maximum eccentricity, i.e.
\begin{align}
    \ev{\rd{\ln a}{t}} &= -\frac{1}{t_{\rm GW, 0}}
            \ev{\frac{1 + 73e^2/24 + 37e^4/96}
                {j^7(e)}}\nonumber\\
        &\equiv -\frac{425/96}{t_{\rm GW, 0}}\frac{1}{j^6(e_{\rm eff})},
        \label{eq:def_e_eff}
\end{align}
where the angle brackets denote averaging over many $t_{\rm ZLK,
oct}$ in order to capture the characteristic eccentricity behavior over many
octupole cycles. In the second line of Eq.~\eqref{eq:def_e_eff}, we have
essentially replaced the ZLK-averaged orbital decay rate by $\rdil{(\ln a)}{t}$
evaluated at $e_{\rm eff}$ multiplied by $j(e_{\rm eff})$. In practice (see
Figs.~\ref{fig:composite_dist}--\ref{fig:composite_bindist}), we typically
average over $2000 t_{\rm ZLK}$ of the non-dissipative simulations to compute
$e_{\rm eff}$.

With $e_{\rm eff}$ computed using Eq.~\eqref{eq:def_e_eff}, we can define the
critical effective eccentricity $e_{\rm eff, c}$ such that the ZLK-averaged
inspiral time is a Hubble time, i.e.\ $\ev{\rdil{(\ln a)}{t}} \equiv
-\p{10\;\mathrm{Gyr}}^{-1}$. This gives
\begin{align}
    j^6\p{e_{\rm eff, c}} \equiv \frac{425}{96}\frac{10\;\mathrm{Gyr}}{t_{\rm
        GW, 0}}, \label{eq:def_e_eff_c}
\end{align}
or equivalently
\begin{align}
    1 - e_{\rm eff, c} \approx 10^{-4}
        \p{\frac{m_{12}}{50M_{\odot}}}
        \p{\frac{q / (1 + q)^2}{1/4}}^{1/3}
        \p{\frac{a}{100\;\mathrm{AU}}}^{-4/3}.
\end{align}
Thus, if a system is evolved using the non-dissipative equations of motion and
satisfies $e_{\rm eff} > e_{\rm eff, c}$, then it is expected to successfully
undergo a smooth merger within a Hubble time.

Therefore, a system can be predicted to merge successfully if it satisfies either
the one-shot or smooth merger criteria. The semi-analytical
merger probability (as a function of $I_0$ and other parameters) is:
\begin{align}
    P_{\rm merger}^{\rm an}\p{I_0; q, e_{\rm out}} =
        P\p{e_{\rm eff} > e_{\rm eff, c} \;\;\text{or}\;\;
        e_{\max} > e_{\rm os}}.\label{eq:def_pmerge_sa}
\end{align}
Although not fully analytical (since numerical integrations of non-dissipative
systems are needed to obtain $e_{\rm eff}$ and $e_{\max}$ in general),
Eq.~\eqref{eq:def_pmerge_sa} provides efficient computation of the merger
probability without full numerical integrations including GW radiation.

The top panels of Figs.~\ref{fig:composite_dist}--\ref{fig:composite_bindist}
show $e_{\rm eff}$ and $e_{\max}$, and their critical values, $e_{\rm eff, c}$
and $e_{\rm os}$. Using these, we compute the semi-analytical
merger probability, shown as the thick green lines in the bottom panels of
Figs.~\ref{fig:composite_dist}--\ref{fig:composite_bindist}. We generally
observe good agreement with the numerical $P_{\rm merger}$. However, $P_{\rm
merger}^{\rm an}$ slightly but systematically underpredicts $P_{\rm merger}$ for
some configurations, such as the prograde inclinations in
Figs.~\ref{fig:composite_1p3} and~\ref{fig:composite_bindist}. These regions
coincide with the inclinations for which the merger outcome is uncertain. This
underprediction is due to the restricted integration time of $2000 t_{\rm ZLK}
\approx 3\;\mathrm{Gyr}$ used for the non-dissipative simulations. To illustrate
this, we also calculate $P_{\rm merger}^{\rm an}$ using a shorter integration
time of $500 t_{\rm ZLK}$ for our non-dissipative simulations. The results are
shown as the light green lines in the bottom panels of
Figs.~\ref{fig:composite_dist}--\ref{fig:composite_bindist}, performing visibly
worse. A more detailed discussion of this issue can be found in
Section~\ref{ss:completeness}.

A few observations about Eq.~\eqref{eq:def_pmerge_sa} can be made. First, it
explains why some prograde systems merge probabilistically ($0 < P_{\rm merger}
< 1$): for the prograde inclinations in Fig.~\ref{fig:composite_1p3},
the $e_{\rm eff}$ values scatter widely around $e_{\rm eff,c}$ [or more
precisely, $j(e_{\rm eff})$ scatters around $j(e_{\rm eff,c})$], even for a
given $I_0$, so the detailed merger outcome depends on the initial conditions.
For the prograde inclinations in Fig.~\ref{fig:composite_1p2}, the double-valued
feature in the $e_{\max}$ plot (the top panel) pointed out in
Section~\ref{ss:oct_gen} represents a sub-population of systems that do not
satisfy Eq.~\eqref{eq:def_pmerge_sa}. Second, $e_{\max} > e_{\rm os}$ often
ensures $e_{\rm eff} > e_{\rm eff, c}$ in practice, as the averaging in
Eq.~\eqref{eq:def_e_eff} is heavily weighted towards extreme eccentricities. As
such, $e_{\rm eff} > e_{\rm eff, c}$ alone is often a sufficient condition in
Eq.~\eqref{eq:def_pmerge_sa}.

The one-shot merger criterion ($e_{\max} > e_{\rm os}$) can also be used to
distinguish two different types of system architectures: if $e_{\lim} \gtrsim
e_{\rm os}$ for a particular architecture, then all initial conditions leading
to orbit flips (i.e., in an octupole-active window) also execute one-shot
mergers. For $e_{\lim} \approx 1$, Eq.~\eqref{eq:def_elim} reduces to
\begin{align}
    j(e_{\lim}) \approx \frac{8\epsilon_{\rm GR}}{9}\p{1 +
        \frac{\eta^2}{12}}^{-1}.
\end{align}
which lets us rewrite the constraint $e_{\lim} \gtrsim e_{\rm os}$ as
\begin{align}
    \p{\frac{a}{a_{\rm out, eff}}} \gtrsim{}&
        0.0186
        \p{\frac{a_{\rm out, eff}}{3600\;\mathrm{AU}}}^{-7/37}
        \p{\frac{m_{12}}{50M_{\odot}}}^{17/37}\nonumber\\
        &\times\p{\frac{30M_{\odot}}{m_3}}^{10/37}
        \p{\frac{q / (1 + q)^2}{1/4}}^{-2/37}.\label{eq:q_237}
\end{align}
For the system architecture considered in
Figs.~\ref{fig:composite_dist}--\ref{fig:composite_e91p5}, this condition is
satisfied, and we see indeed that wherever the top panel suggests orbit flipping
($e_{\max} = e_{\lim}$), the bottom panel shows $P_{\rm merger} \approx 1$. When
the condition (Eq.~\ref{eq:q_237}) is not satisfied, one-shot mergers are not
possible, and $P_{\rm merger}$ is generally only nonzero for a small range about
$I_{\rm 0, \lim}$.

\section{Merger Fraction as a Function of Mass Ratio}\label{s:merger_frac}

Having developed an semi-analytical understanding of the binary
merger window and probability in the last section (particularly
Section~\ref{ss:nogw_merger}), we now study the fraction of BH binaries in
triples that successfully merge under various conditions -- we 
call this the merger fraction.

\subsection{Merger Fraction for Fixed Tertiary Eccentricity}

\begin{figure*}
    \centering
    \includegraphics[width=\textwidth]{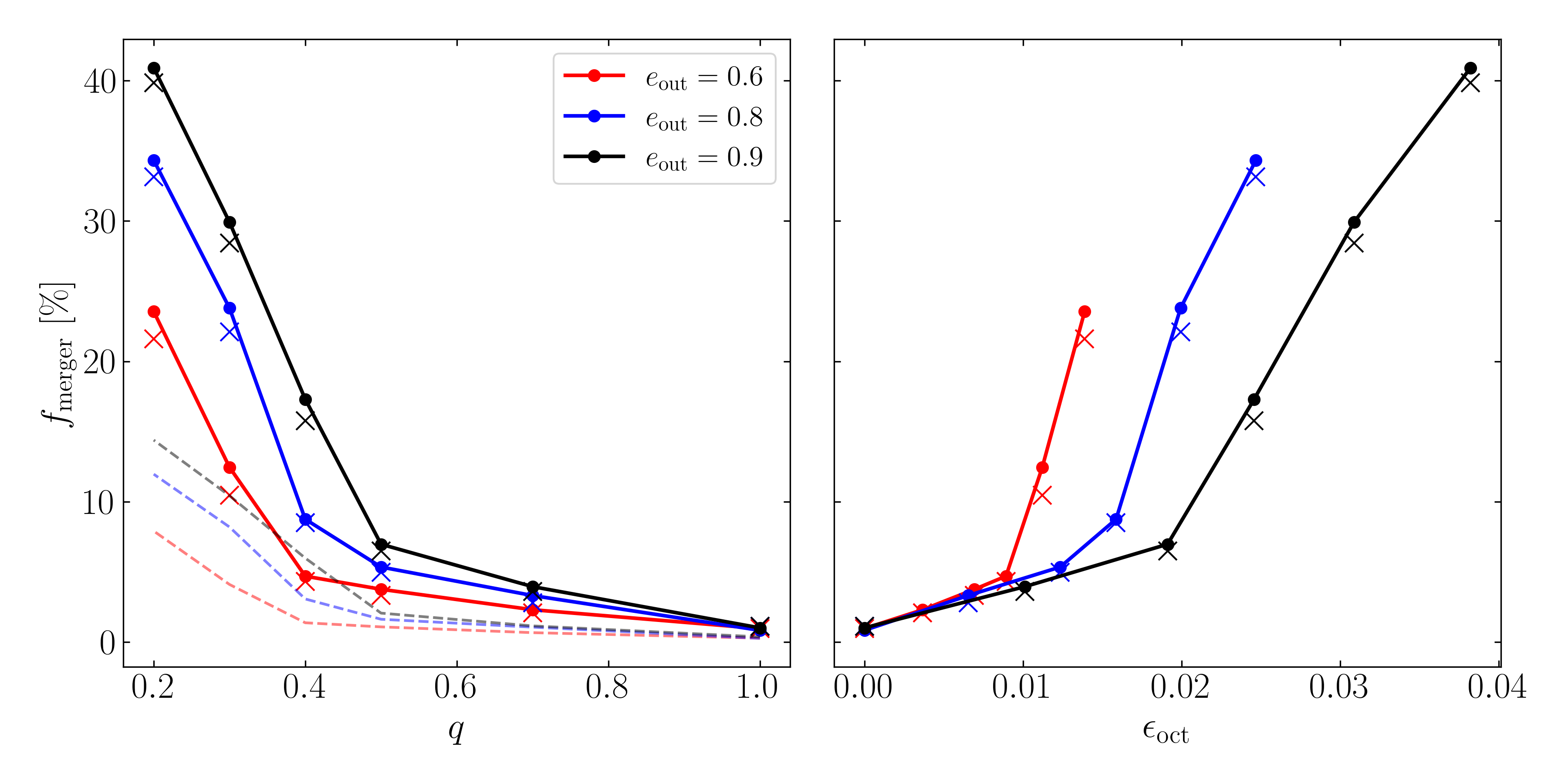}
    \caption{Merger fraction (Eq.~\ref{eq:def_pmerge}) of BH binaries in
    triples as a function of mass ratio $q$ (left panel) for several values of
    outer binary eccentricities. The other system parameters are the same as in
    Figs.~\ref{fig:composite_dist}--\ref{fig:composite_e91p5}. The right panel
    shows the same merger fraction, but plotted against the octupole parameter
    $\epsilon_{\rm oct}$.  The filled circles joined by the solid lines are
    numerical results (based on integrations for full triple system evolution
    including GW emission; see the black solid lines in the bottom panels of
    Figs.~\ref{fig:composite_dist}--\ref{fig:composite_e91p5})
    assuming random mutual inclinations between the inner and
    outer binaries (uniform in $\cos I_0$), and the dashed lines denote the
    merger fractions if the mutual inclinations are distributed according to
    Eq.~\eqref{eq:P_wedge}. The crosses are semi-analytical
    results using an integration time of $2000 t_{\rm ZLK}$ (see the thick green
    lines in the bottom panels of
    Figs.~\ref{fig:composite_dist}--\ref{fig:composite_e91p5}).
    }\label{fig:total_merger_fracs}
\end{figure*}

We first consider the simple case where $e_{\rm out}$ is fixed at a few
specific values and compute the merger fraction as a function of the mass ratio
$q$. We consider isotropic mutual orientations between the inner and outer
binaries, i.e.\ we draw $\cos I_0$ from a uniform grid over the
range $[-1, 1]$ (recall that $\omega$, $\omega_{\rm out}$, and $\Omega$ are
drawn uniformly from the range $[0, 2\pi)$ 
when computing the merger probability $P_{\rm merger}$ at a given $I_0$). The
merger fraction is then given by:
\begin{align}
    f_{\rm merger}\p{q, e_{\rm out}} \equiv
        \frac{1}{2}\int\limits_{-1}^1\mathrm{d}\cos I_0\;
            P_{\rm merger}\p{I_0; q, e_{\rm out}} .\label{eq:def_fmerge}
\end{align}
This is proportional to the integral of the black lines (weighted by $\sin I_0$)
in the bottom panels of
Figs.~\ref{fig:composite_dist}--\ref{fig:composite_e91p5}. We can also use
semi-analytical criteria introduced in
Section~\ref{ss:nogw_merger} to predict the outcome and merger fraction. This is
computed by using $P_{\rm merger}^{\rm an}$ as the integrand in
Eq.~\eqref{eq:def_fmerge}, or by evaluating the integral of the thick green
lines (weighted by $\sin I_0$) in the bottom panels of
Figs.~\ref{fig:composite_dist}--\ref{fig:composite_e91p5}.
Figure~\ref{fig:total_merger_fracs} shows the resulting $f_{\rm merger}$ and the
analytical estimates for all combinations of $q \in \z{0.2, 0.3, 0.4, 0.5, 0.7,
1.0}$ and $e_{\rm out} \in \z{0.6, 0.8, 0.9}$. It is clear that the numerical
$f_{\rm merger}$ and the analytical estimate agree well, and that the merger
fraction increases steeply for smaller $q$.

To explore the impact of our choice of isotropic mutual
orientations between the two binaries, we also consider a wedge-shaped
distribution of $\cos I_0$ as was found in the population synthesis studies of
\citet{antonini2017binary}. We still use the same uniform grid of $\cos I_0$ as
before, but weight each eccentricity by its probability probability density
following the distribution:
\begin{equation}
    P\p{\cos I_0} = \frac{1}{4} + \frac{\abs{\cos I_0}}{2}.\label{eq:P_wedge}
\end{equation}
The resulting $f_{\rm merger}$ for a tertiary with $\cos I_0$ distributed like
Eq.~\eqref{eq:P_wedge} is shown as the dashed lines in
Fig.~\ref{fig:total_merger_fracs}. While the total merger fractions decrease,
the strong enhancement of the merger fraction at smaller $q$ is unaffected.

In the right panel of Fig.~\ref{fig:total_merger_fracs}, we see that the merger
fractions for the three $e_{\rm out}$ values overlap for small $\epsilon_{\rm
oct}$. This implies that $f_{\rm merger}$ depends only on $\epsilon_{\rm oct}$
in this regime, and not on the values of $q$ and $e_{\rm out}$ independently.
From Fig.~\ref{fig:composite_dist} (which has $\epsilon_{\rm oct} = 0.007$), we
see that this suggests that the size of the retrograde merger window only
depends on $\epsilon_{\rm oct}$, much like what Eq.~\eqref{eq:I_oct_MLL} shows
for the test-particle limit. However, once $\epsilon_{\rm oct}$ is increased
sufficiently, the three curves in the right panel of
Fig.~\ref{fig:total_merger_fracs} cease to overlap. This can be attributed to
their different $\eta$ values: for sufficiently small $\epsilon_{\rm oct}$, no
prograde initial inclinations successfully merge (e.g.,
Fig.~\ref{fig:composite_dist}), and the merger fraction is solely determined by
the size of the retrograde octupole-active window. But once $\epsilon_{\rm oct}$
is sufficiently large, prograde mergers become possible, and the merger fraction
is also affected by the size of the octupole-inactive gap, which depends on
$\eta$. This again illustrates the importance of the octupole-inactive gap,
which we comment on in Appendix~\ref{app:gap}.

Figure~\ref{fig:sweepbin_simpleouter} depicts the merger fractions for systems
with $a_0 = 50\;\mathrm{AU}$ (the other parameters are the same as in
Fig.~\ref{fig:total_merger_fracs}). According to Eq.~\eqref{eq:q_237}, these
systems no longer satisfy $e_{\lim} \gtrsim e_{\rm os}$, so the
merger fraction is expected to diminish strongly and vary much more weakly with
$q$, as one-shot mergers are no longer possible. This is indeed observed,
particularly for the $e_{\rm out} = 0.6$ curve in
Fig.~\ref{fig:sweepbin_simpleouter}. We also remark that the
semi-analytical prediction accuracy is poorer in this case than in
Fig.~\ref{fig:total_merger_fracs}. This is because the only mergers in this
regime are smooth mergers. As can be seen for the prograde $I_0$ in
Figs.~\ref{fig:composite_1p3} and~\ref{fig:composite_bindist}, smooth mergers
occur over a wide range of merger times $T_{\rm m}$, and the specific $T_{\rm
m}$ that a system experiences depends sensitively on its chaotic evolution.
Thus, Eq.~\eqref{eq:def_e_eff} is a rather approximate estimate of the amount of
GW emission that a real system emits during a smooth merger; indeed, the
prograde regions of Figs.~\ref{fig:composite_1p3}
and~\ref{fig:composite_bindist} show that the merger times for smooth mergers
are systematically underpredicted by the semi-analytic merger criterion (see
discussion in Section~\ref{ss:completeness}). The non-monotonicity of the
semi-analytic merger fraction for $e_{\rm out} = 0.6$ from $q = 0.2$ to $q =
0.3$ is due to small sample sizes and finite grid spacing in $\cos I_0$.

\begin{figure*}
    \centering
    \includegraphics[width=\textwidth]{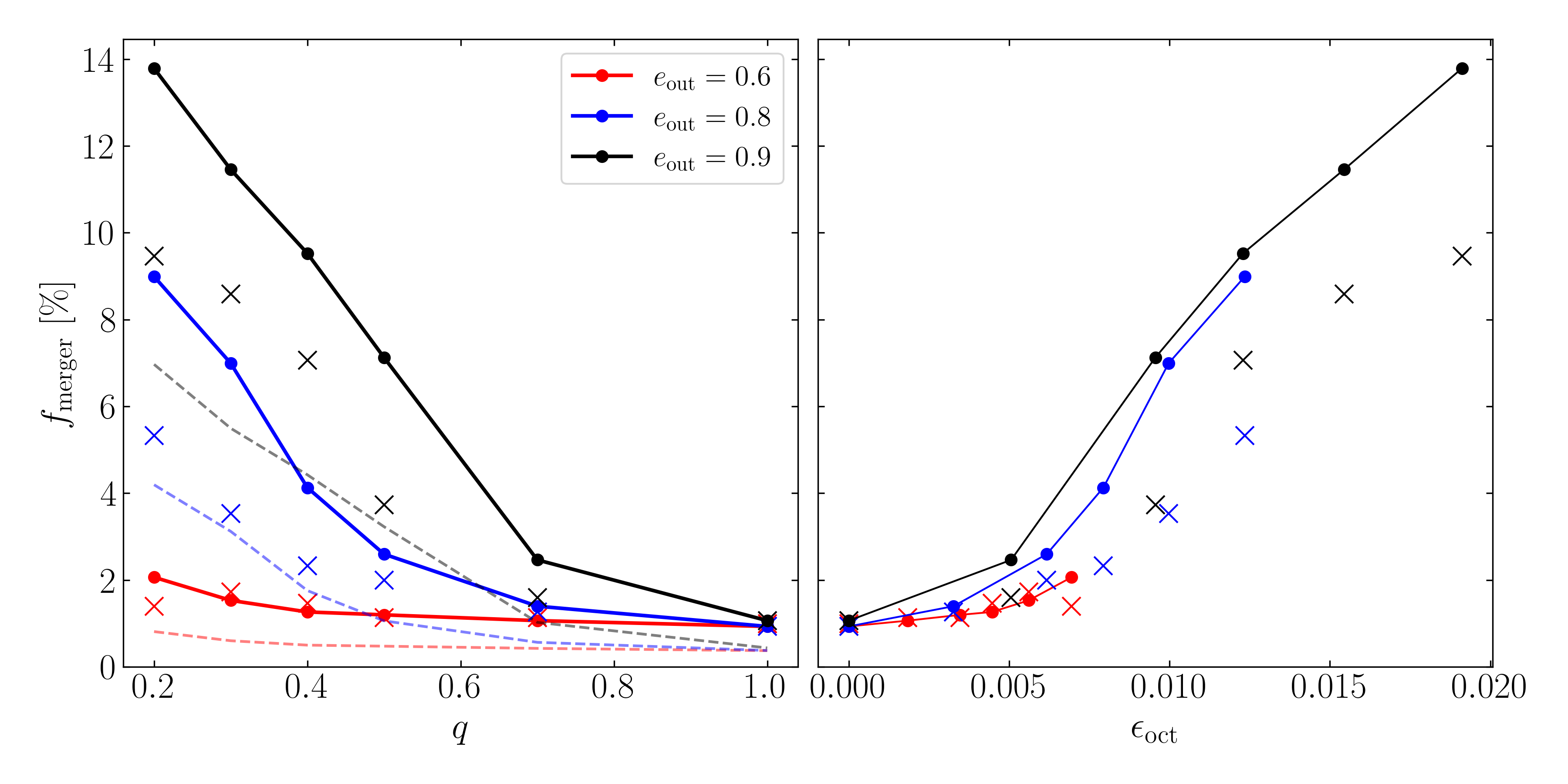}
    \caption{Same as Fig.~\ref{fig:total_merger_fracs} but for $a_0 =
    50\;\mathrm{AU}$.}\label{fig:sweepbin_simpleouter}
\end{figure*}

\subsection{Merger Fraction for a Distribution of Tertiary Eccentricities}

For a distribution of tertiary eccentricities, denoted $P\p{e_{\rm out}}$, the
merger fraction is given by
\begin{align}
    \eta_{\rm merger}(q) &= \int \mathrm{d}e_{\rm out}\;
            P\p{e_{\rm out}} f_{\rm merger}\p{q, e_{\rm out}},\nonumber\\
        &= \int\limits\mathrm{d} e_{\rm out}\; \frac{P\p{e_{\rm out}}}{2}
            \int\limits_{-1}^1\mathrm{d} \cos I_0\; P_{\rm merger}\p{I_0; q,
            e_{\rm out}}.
        \label{eq:def_eta_merge}
\end{align}
We consider two possible $P(e_{\rm out})$ with $e_{\rm out} \in [0, 0.9]$: (i) a
uniform distribution, $P\p{e_{\rm out}} = \textrm{constant}$, and (ii) a thermal
distribution, $P\p{e_{\rm out}} \propto e_{\rm out}$.

\begin{figure}
    \centering
    \includegraphics[width=\columnwidth]{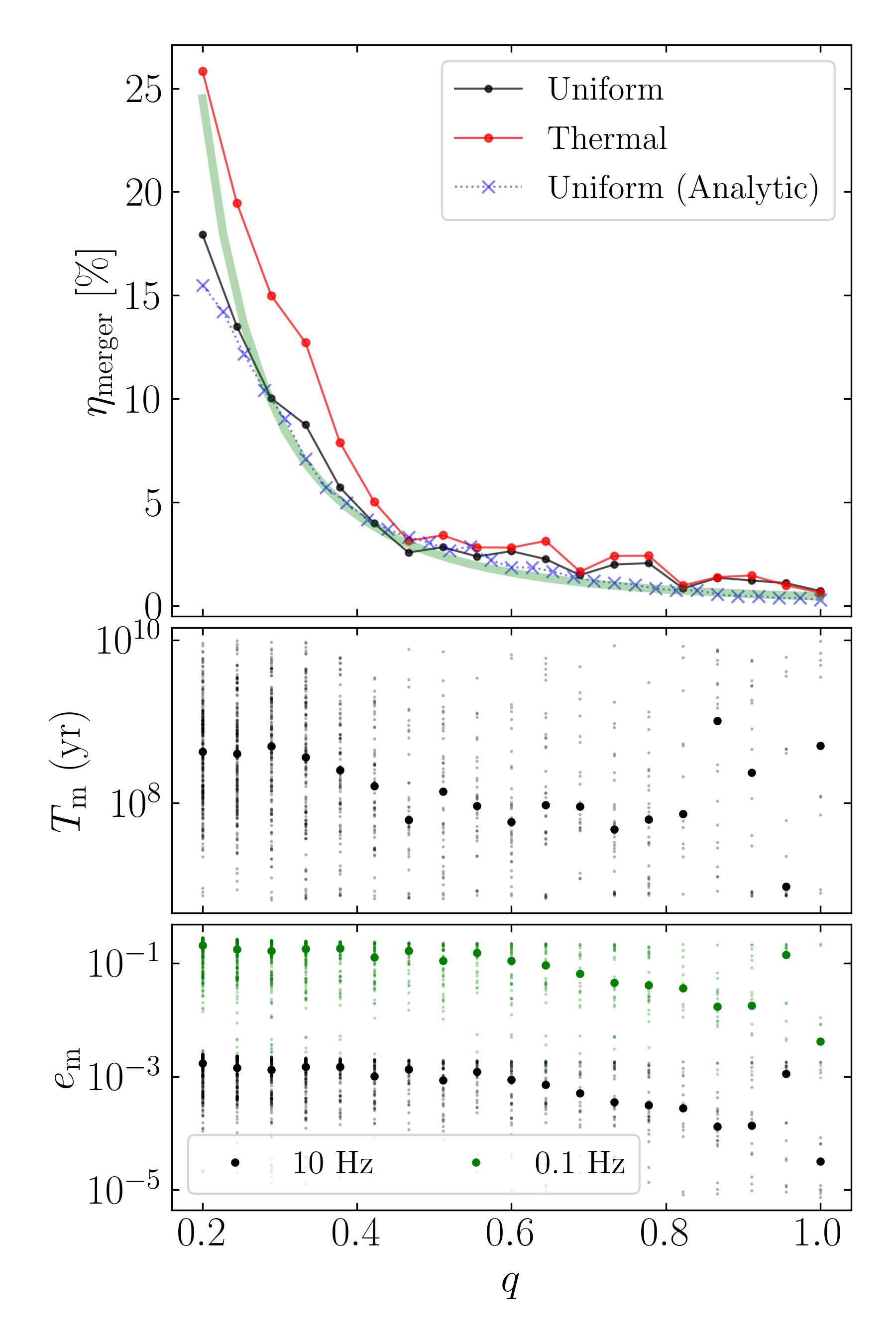}
    \caption{\emph{Upper panel:} Binary BH merger fraction as a function of mass
    ratio $q$ for the fiducial triple systems (with parameters the same as in
    Figs.~\ref{fig:composite_dist}--\ref{fig:composite_e91p5}), assuming random
    mutual inclinations (uniform in $\cos I_0$), and either uniform (black dots)
    or thermal distribution (red dots) for the tertiary eccentricity
    distribution [with $e_{\rm out} \in [0,0.9]$]. These are obtained
    numerically using Eq.~\eqref{eq:def_eta_merge} by sampling $1000$
    combinations of $e_{\rm out}$, $\cos I_0$, $\omega$, $\omega_{\rm out}$, and
    $\Omega$. The blue dotted line is the semi-analytical
    result obtained by applying Eq.~\eqref{eq:def_pmerge_sa} in
    Eq.~\eqref{eq:def_eta_merge} (evaluated using a dense uniform grid of $\cos
    I_0$ and $e_{\rm out}$). The thick green line is a power-law fit to the
    analytical $\eta_{\rm merger}$ with a power law index of $-2.5$.
    \emph{Middle panel:} Merger times of successful mergers for a uniform
    $e_{\rm out}$ distribution (the median is denoted with the large black dot).
    \emph{Bottom panel:} Merging binary eccentricities (again, for a uniform
    $e_{\rm out}$ distribution) in the LISA band ($0.1\;\mathrm{Hz}$; green) and
    in the LIGO band ($10 \;\mathrm{Hz}$; black), with medians marked with large
    dots. }\label{fig:popsynth}
\end{figure}

The top panel of Fig.~\ref{fig:popsynth} shows $\eta_{\rm merger}$ (black
dots) for the fiducial triple systems (with the same parameters as in
Figs.~\ref{fig:composite_dist}--\ref{fig:composite_e91p5}). For each $q$, the
integral in Eq.~\eqref{eq:def_eta_merge} is computed using $1000$ realizations
of random $e_{\rm out}$, $\cos I_0$, $\omega$, $\omega_{\rm out}$, and
$\Omega$. Not surprisingly, we see $\eta_{\rm merger}$ increases with decreasing
$q$. When $q$ is small, a thermal distribution of $e_{\rm out}$ tends to yield
higher $\eta_{\rm merger}$ than does a uniform distribution. We also compute the
merger fraction using the semi-analytical merger probability of
Eq.~\eqref{eq:def_pmerge_sa} on a dense grid of initial conditions uniformly
sampled in $e_{\rm out}$ and $\cos I_0$; the result is shown as the blue dotted
line in Fig.~\ref{fig:popsynth}, which is in good agreement with the
uniform-$e_{\rm out}$ simulation result (black).

To characterize the properties of merging binaries, the middle and bottom panels
of Fig.~\ref{fig:popsynth} show the distributions of merger times and merger
eccentricities (at both the LISA and LIGO bands) for different mass ratios. To
obtain the LISA and LIGO band eccentricities (with GW frequency equal to
$0.1\;\mathrm{Hz}$ and $10\;\mathrm{Hz}$ respectively), the inner binaries are
evolved from when they reach $0.005 a_0$ (at which point we terminate the
integration of the triple system evolution as the inner binary's
evolution is decoupled from the tertiary; see Eq.~\ref{eq:epsgr_0005}) to
physical merger using Eqs.~(\ref{eq:def_tgw}--\ref{eq:dedt_gw}). While the LIGO
band eccentricities are all quite small ($\lesssim 10^{-3}$), the LISA band
eccentricities (at $0.1 \;\mathrm{Hz}$) are significant, with median $\gtrsim
0.2$ for $q\lesssim 0.5$. We note that these eccentricities are generally
smaller than those found in the population studies of \citet{LL19}. This is
because in this paper we consider only sufficiently hierarchical systems for
which double-averaged evolution equations are valid, whereas \citet{LL19}
included a wider range of triple hierarchies and had to use $N$-body
integrations to evolve some of the systems.

\begin{figure}
    \centering
    \includegraphics[width=\columnwidth]{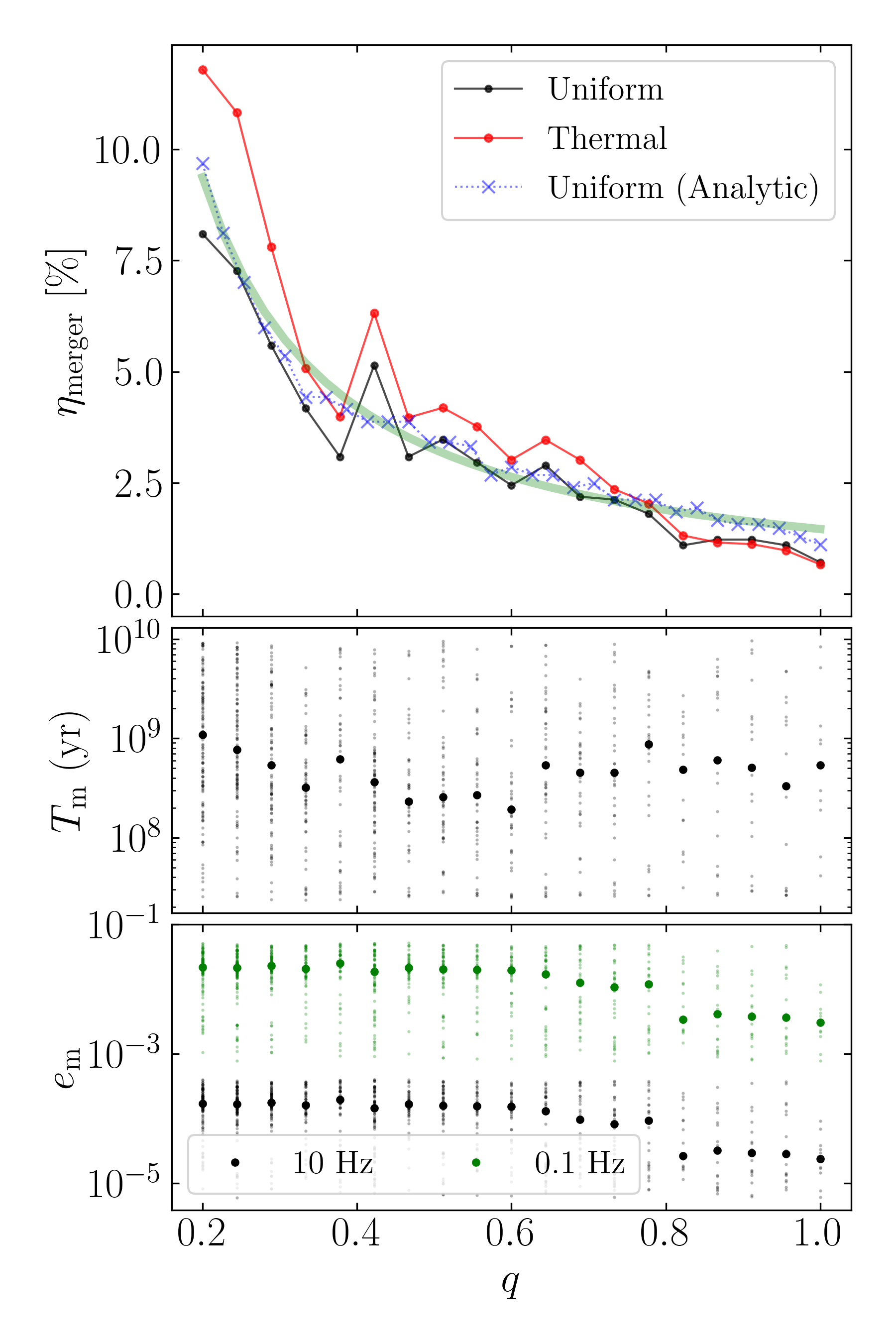}
    \caption{Same as Fig.~\ref{fig:popsynth} but for $a_{\rm out, eff} =
    5500\;\mathrm{AU}$. The power law index of the fit to the
    analytical $\eta_{\rm merger}$ is $-1.15$. }\label{fig:popsynth5500}
\end{figure}

For comparison, Figure~\ref{fig:popsynth5500} shows the results when
$a_{\rm out, eff} = 5500\;\mathrm{AU}$ (instead of $a_{\rm out, eff} =
3600\;\mathrm{AU}$ for Fig.~\ref{fig:popsynth}) with all other parameters
unchanged. While $\eta_{\rm merger}$ is lower than it is for $a_{\rm out, eff} =
3600\;\mathrm{AU}$, there is still a large increase of $\eta_{\rm merger}$ with
decreasing $q$. Since Eq.~\eqref{eq:q_237} is still satisfied, this is expected.

\subsection{$q \ll 1$ Limit}

For fixed $m_{12}$ (and other parameters), even though the octupole strength
$\epsilon_{\rm oct}$ increases as $q$ decreases, the efficiency of GW
radiation also decreases. It is therefore natural to ask at what $q$ these
competing effects become comparable and the merger fraction is maximized. We
show that this does not happen until $q$ is extremely small.

\begin{figure}
    \centering
    \includegraphics[width=\columnwidth]{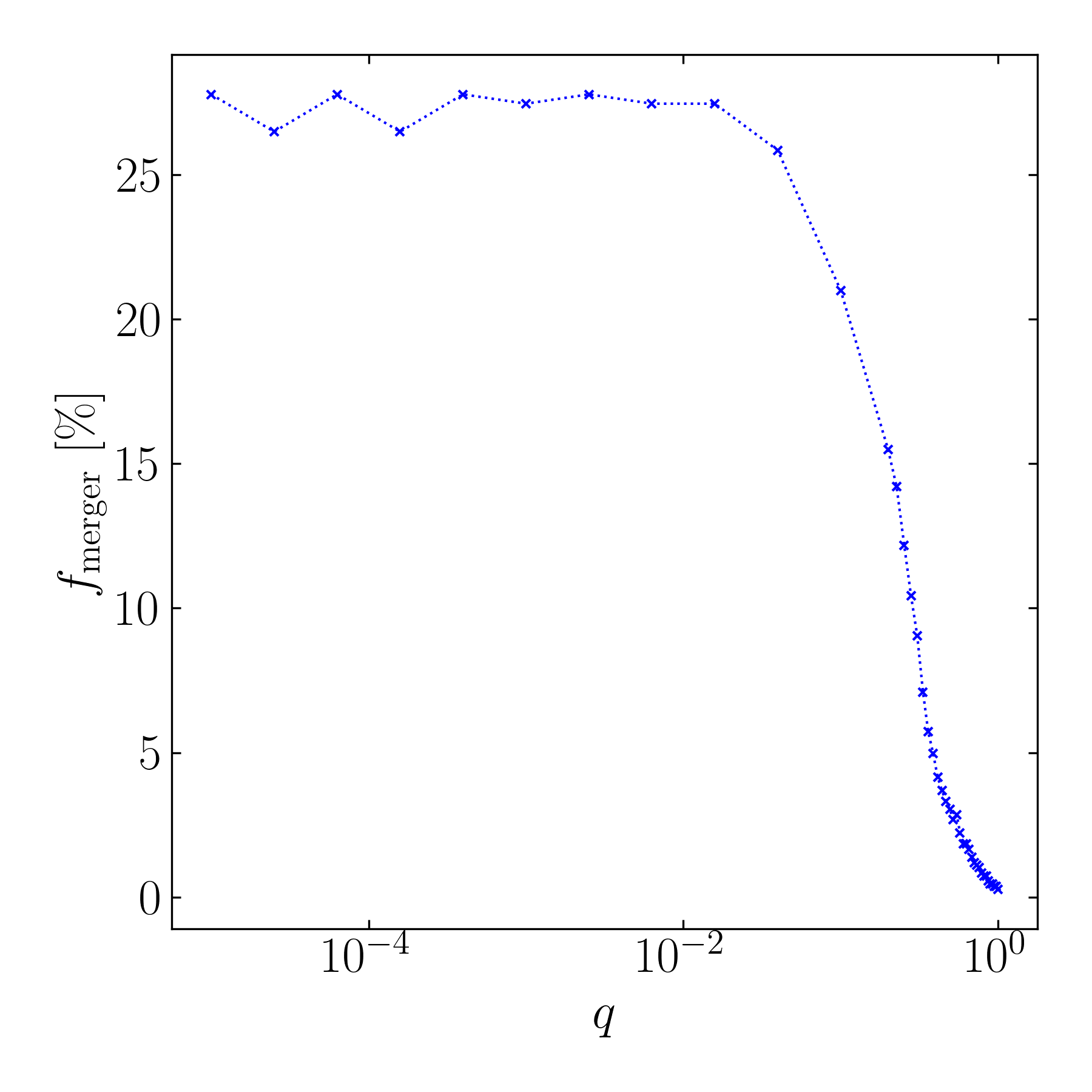}
    \caption{Same as blue dashed line of the top panel of
    Fig.~\ref{fig:popsynth} but extended to very small $q$. Due to the very weak
    $q$ dependence in Eq.~\eqref{eq:q_237}, $f_{\rm merger}$ is expected to
    depend very weakly on $q$ when $q \ll 1$ (such that $\epsilon_{\rm oct}$ is
    approximately constant), which agrees with the simulation
    results.}\label{fig:popsynth_lowq}
\end{figure}

We see from Figs.~\ref{fig:composite_dist}--\ref{fig:composite_e91p5} that
$e_{\lim} > e_{\rm os}$ for our fiducial triple systems. Indeed,
from Eq.~\eqref{eq:q_237}, we see that even for $q$ as small as $10^{-5}$, the
condition $e_{\lim} > e_{\rm os}$ is satisfied. This implies that most binaties
execute one-shot mergers when undergoing an orbit flip. In addition, recall that
the characteristic time for the binary to approach $e_{\lim}$ can be estimated
by Eq.~\eqref{eq:def_tzlkoct}, which, for our fiducial triple systems, is given
by
\begin{align}
    t_{\rm ZLK, oct} \simeq{}& 10^8
        \p{\frac{m_{12}}{50M_{\odot}}}^{1/2}
        \p{\frac{a_{\rm out, eff}}{3600\;\mathrm{AU}}}^{7/2}
        \p{\frac{a}{100\;\mathrm{AU}}}^{-2}\nonumber\\
        &\times \p{\frac{m_3}{30M_{\odot}}}^{-1}
            \Bigg[\frac{1 - q}{1 + q}\frac{e_{\rm out}}{\sqrt{1 - e_{\rm out}^2}}
            \Bigg]^{-1/2}
            \;\mathrm{yr}.
\end{align}
Since $t_{\rm ZLK, oct } \ll 10\;\mathrm{Gyr}$, this implies that the
octupole-ZLK-induced binary merger fractions are primarily determined by what
initial conditions would lead to extreme eccentricity excitation and only weakly
depend on the GW radiation rate. Indeed, Eq.~\eqref{eq:q_237} shows that, while
$e_{\lim} > e_{\rm os}$ is indeed violated if $q$ is decreased sufficiently, the
dependence is extremely weak. Thus, $\eta_{\rm merger}$ is expected to be very
nearly constant for all physically relevant values of $q$, as can be seen in
Fig.~\ref{fig:popsynth_lowq}.

\subsection{Limitations of semi-analytic
Calculation}\label{ss:completeness}

It can be seen in Fig.~\ref{fig:total_merger_fracs} that the
semi-analytical merger fractions are systematically lower than
the values obtained from the direct simulations. One reason that this
discrepancy arises is because the non-dissipative simulations used to compute
$e_{\rm eff}$ and $e_{\max}$ are only run for $2000 t_{\rm LK} \approx
3\;\mathrm{Gyr}$, while the full simulations including GW dissipation are run
for $10\;\mathrm{Gyr}$. Owing to the chaotic nature of the octupole-order ZLK
effect, this means that, if an initial condition leads to extreme eccentricities
only after many Gyrs, then $e_{\rm eff}$ and $e_{\max}$ are underpredicted by
the non-dissipative simulations. Additionally, there are times when eccentricity
vector of the inner binary is librating, during which orbit flips are strongly
suppressed \citep{katz2011long}. Since the librating phase can last an
unpredictable amount of time, this suggests that the
semi-analytical merger criteria can become more complete as the
integration time is increased.

\begin{figure}
    \centering
    \includegraphics[width=\columnwidth]{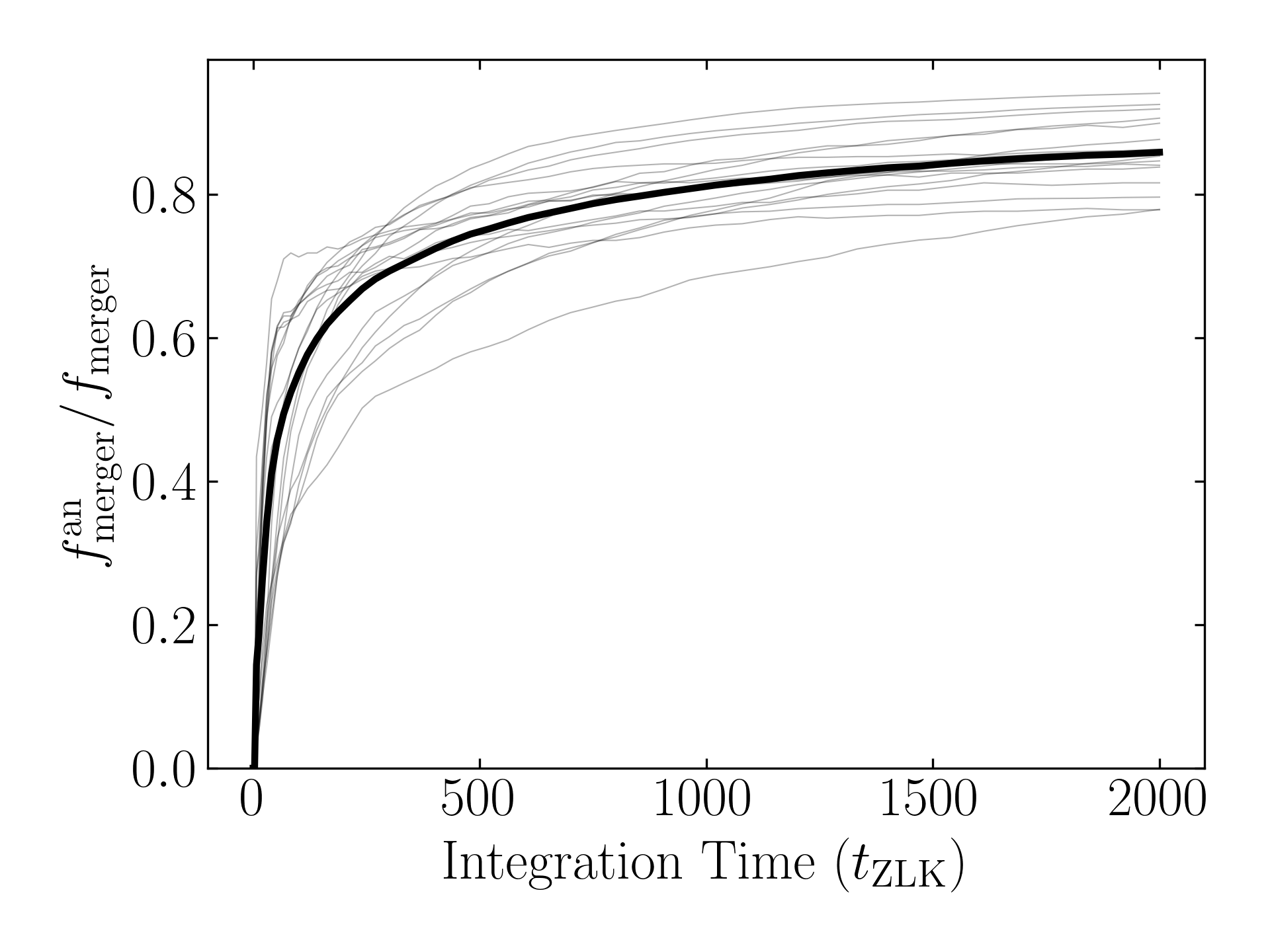}
    \caption{Completeness of the semi-analytical merger fraction, defined as $f_{\rm
    merger}^{\rm an} / f_{\rm merger}$, as a function of the integration time
    used for the non-dissipative simulations, in the fiducial parameter regime
    while $e_{\rm out}$ is fixed at a few values. The thin grey lines indicate
    the completeness for particular combinations of $(q, e_{\rm out})$, and the
    thick black line denotes their average. We see that completeness is still
    increasing as the integration time approaches $2000 t_{\rm ZLK} \approx
    3\;\mathrm{Gyr}$. }\label{fig:completeness}
\end{figure}

We quantify the ``completeness'' of the semi-analytical merger
fraction via the ratio $f_{\rm merger}^{\rm an} / f_{\rm merger}$ as a function
of non-dissipative integration time. We focus on the fiducial triple systems for
demonstrative purposes and compute the completeness for each of the $q$ and
$e_{\rm out}$ combinations shown in Fig.~\ref{fig:total_merger_fracs}.
Figure~\ref{fig:completeness} shows the completeness for each of these
simulations in light grey lines and their mean in the thick black line. We see
that the completeness is still increasing even as the non-dissipative simulation
time is increased to $2000t_{\rm ZLK}$, so we expect that even longer
integration times would give even better agreement with the dissipative
simulations.

\section{Mass Ratio Distribution of Merging BH Binaries}\label{s:q_dist}

In Section 4, we have calculated the binary BH merger fractions $f_{\rm merger}$
and $\eta_{\rm merger}$ as a function of the mass ratio $q$ for some
representative triple systems. To determine the distribution in $q$ and $m_{12}$
(total mass) of the merging binaries, we would need to know both the
initial distribution in $q$, $m_{12}$ and $a_0$ of the inner BH binaries and the
distribution in $m_3$, $a_{\rm out}$ and $e_{\rm out}$ of the outer
binaries, denoted by:
\begin{align}
    \frac{\md F}{ \md q \md m_{12} \md a_0},\quad
    \frac{\md F_{\rm out}}{ \md m_3 \md a_{\rm out} \md e_{\rm out}}.
\end{align}
The distribution in $q$ and $m_{12}$ of the merging binaries is then
\begin{align}
    & {\md F_{\rm merger}\over \md q \md m_{12}} = \int \!\md a_0 \md m_{\rm
        out}\md a_{\rm out}\md e_{\rm out}\, {\md F\over \md q \md m_{12} \md
        a_0}\nonumber\\
    &\qquad \times {\md F_{\rm out}\over \md m_3 \md a_{\rm out} \md  e_{\rm out}}\,
      f_{\rm merger}(q,e_{\rm out}; m_{12},a_0,m_3,a_{\rm out},e_{\rm out}),
      \label{eq:dong1}
\end{align}
where $f_{\rm merger}$ is given by Eq.~\eqref{eq:def_fmerge} (assuming random
mutual inclinations between the inner and outer binaries), and we have spelled
out its dependence on various system parameters. Some examples of $f_{\rm
merger}$ are shown in
Figs.~\ref{fig:total_merger_fracs}--\ref{fig:sweepbin_simpleouter}. If we
further specify the eccentricity distribution of the outer binaries, we have
\begin{align}
    & {\md F_{\rm merger}\over \md q \md m_{12}} = \int
        \!\md a_0 \md m_3\md a_{\rm out}\,
        {\md F\over \md q \md m_{12} \md a_0}\nonumber\\
    &\qquad \times {\md F_{\rm out}\over \md m_3 \md a_{\rm out,eff}}\,
        \eta_{\rm merger}(q; m_{12},a_0,m_3,a_{\rm out,eff}),
        \label{eq:dong2}
\end{align}
where $\eta_{\rm merger}$ is given by Eq.~\eqref{eq:def_eta_merge}. Some
examples of $\eta_{\rm merger}$ are shown in the top panels of
Figs.~\ref{fig:popsynth}--\ref{fig:popsynth5500}.

Clearly, to properly evaluate Eq.~\eqref{eq:dong1} or~\eqref{eq:dong2} would
require large population synthesis calculations and in any case would involve
significant uncertainties, a task beyond the scope of this paper. For
illustrative purposes, we consider the fiducial triple systems as studied in
Section~\ref{s:merger_frac}, and estimate the mass-ratio distribution of BH
mergers as
\begin{align}
   {\md F_{\rm merger}\over \md q}\sim {\md F\over \md q}\, \eta_{\rm merger}(q).
    \label{eq:gmerge_dq}
\end{align}

\subsection{Initial $q$-distribution of BH Binaries}\label{ss:qdist_init}

The initial mass-ratio distribution of BH binaries, $\md F/\md q$, is uncertain.
It can be derived from the the mass distributions of of main-sequence (MS)
binaries, together with the MS mass ($m_{\rm ms}$) to BH mass ($m$) relation.

For the distribution of MS binary masses, we assume that each MS component mass
is drawn from a Salpeter-like initial mass function (IMF) independently, with
\begin{align}
    {\md F_{\rm ms} \over \md m_{\rm ms} }\propto m_{\rm ms}^{-\alpha},
        \label{eq:def_alpha}
\end{align}
in the range $m_{\min}\le m_{\rm ms} \leq m_{\max}$. Note in this case the MS
binary mass-ratio distribution is (for $q\leq 1$)
\begin{align}
    {\md F_{\rm ms}\over \md q}\propto q^{\alpha -2}
        \left[ 1-\left( \frac{q}{q_{\min}}\right)^{2-2\alpha}\right],
\end{align}
where $q_{\min} = m_{\min}/m_{\max}$ is the minimum possible binary mass ratio
\citep[this is a generalization of the result of][]{tout_dist}. We consider two
representative values of $\alpha$: (i) $\alpha = 2.35$, the canonical Salpeter
IMF \citep{salpeter1955luminosity}, and (ii) $\alpha = 2$, resulting in a
uniform $q$ distribution (for $q \gtrsim 2q_{\min}$). The latter case is
consistent with observational studies of the mass ratio of high-mass MS binaries
\citep{sana2012binary, duchene2013, kobulnicky2014, moe2017mind}.

To obtain $\rdil{F}{q}$, we compute the BH binary mass ratio when each main
sequence mass $m_{\rm ms}$ is mapped to its corresponding BH mass $m$. This
mapping is taken from \citet{spera2017very} for the mass range $25M_{\odot} \leq
m_{\rm ms} \leq 117 M_{\odot}$. We consider both the case where $Z = 0.02$
(``high $Z$'') and where $Z = 2.0 \times 10^{-4}$ (``low $Z$''), the two
limiting metallicities used in \citet{spera2017very}. We can then numerically
compute $\rdil{F}{q}$ by sampling masses for stellar binaries from the IMF,
translating these into BH masses, then calculating the resulting BH mass ratios
for each binary. The upper panel of Fig.~\ref{fig:qdist_salpeter} shows the
$\rdil{F}{q}$ obtained via this procedure for a Salpeter IMF ($\alpha = 2.35$)
when sampling $10^5$ MS binaries for each metallicity. In the lower four panels,
we also show $\rdil{F}{q}$ restricted to particular ranges of $m_{12}$. Note
that the distributions differ significantly among the $m_{12}$ ranges and also
between the two metallicities. Figure~\ref{fig:qdist_uniform} shows the case
when $\alpha = 2$, which mostly resembles Fig.~\ref{fig:qdist_salpeter}.

\subsection{$q$-distribution of Merging BH Binaries}\label{ss:qdist_merge}

\begin{figure}
    \centering
    \includegraphics[width=0.7\columnwidth]{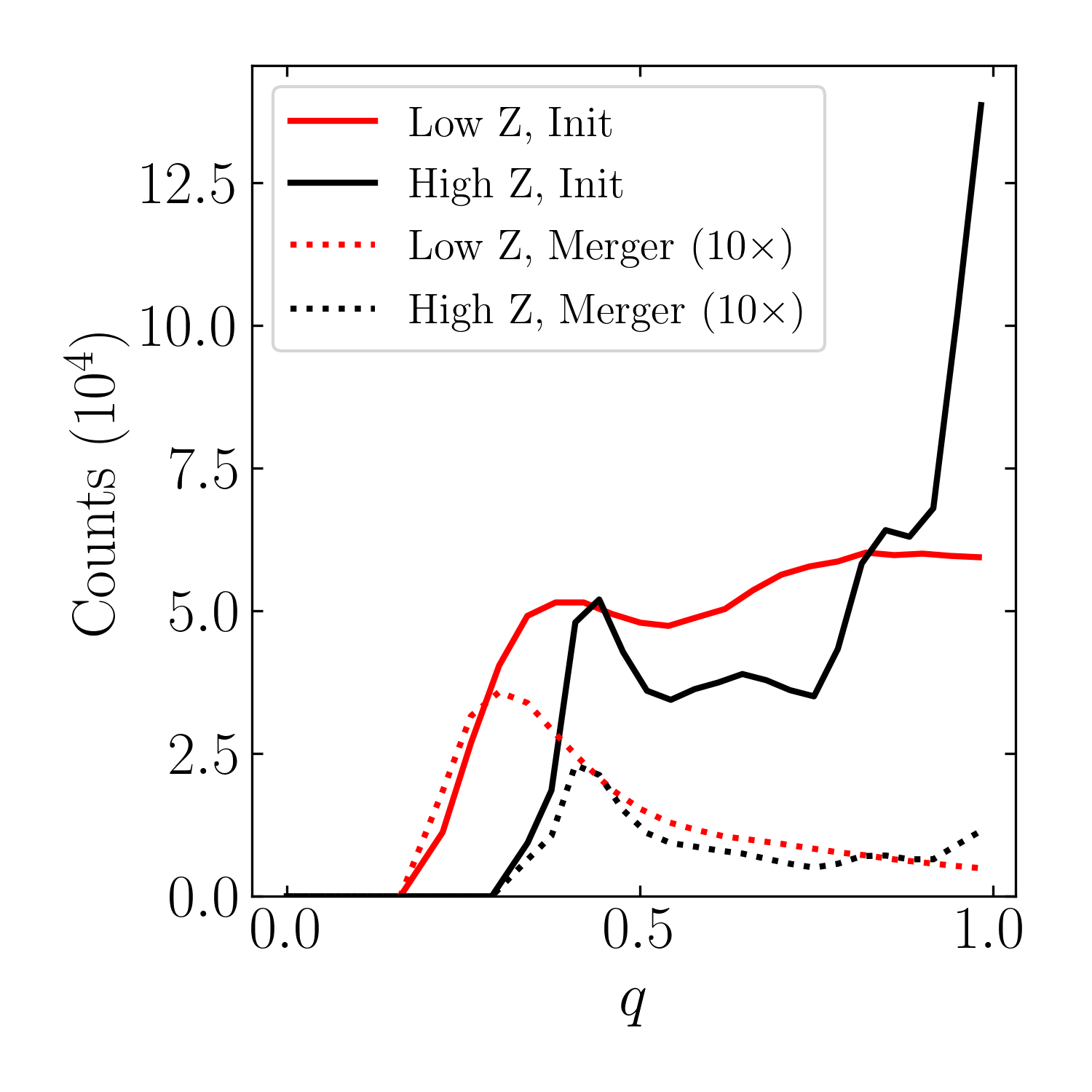}
    \includegraphics[width=\columnwidth]{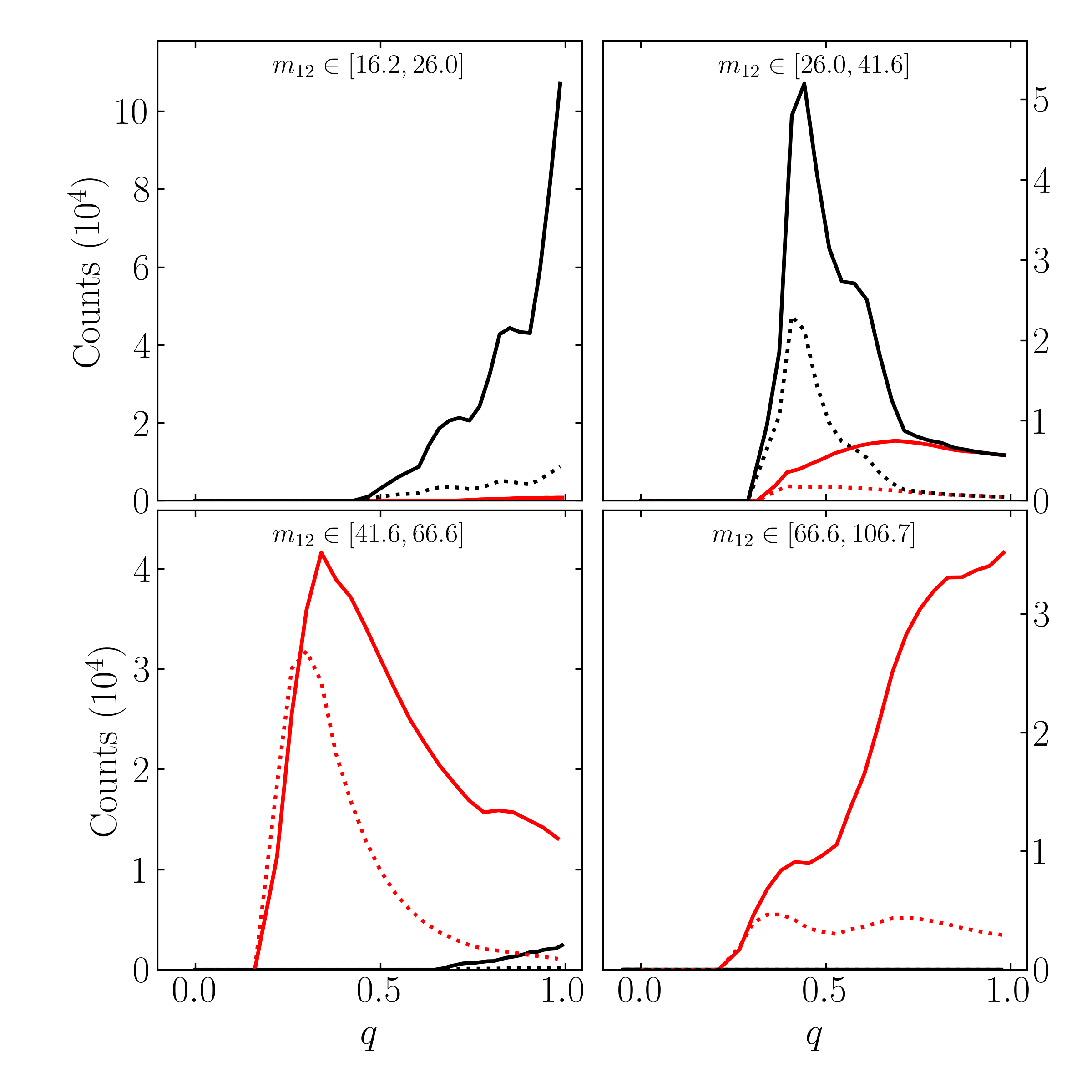}
    \caption{Mass ratio distributions of the initial BH binaries (solid lines)
    and merging BH binaries (dotted lines) when using $\alpha = 2.35$ for the MS
    stellar initial mass function (see Sections~\ref{ss:qdist_init}
    and~\ref{ss:qdist_merge}). \emph{Top panel:} Distribution of binary mass
    ratio at formation and merger for all possible total binary BH masses. Each
    BH mass is obtained from the MS mass using the fitting formula of
    \citet{spera2017very} for metallicities of $2 \times 10^{-4}$ (Low Z) and
    $0.02$ (High Z), while the merger fraction of BH binaries is given by
    Eq.~\eqref{eq:etam_approx}. To produce these distributions, $10^5$ initial
    MS binaries are used for each metallicity, and the number of merging BH
    binaries has been scaled up by a factor of $10$ for visibility. The counts
    refer to the number per $\Delta q = 0.05$ bin. \emph{Bottom four panels:}
    Same as the top panel but with specific ranges of $m_{12}$, the total BH
    mass of the binary (as labeled). Note that low-$m_{12}$ systems are mainly
    produced from high-Z MS binaries, while high-$m_{12}$ systems are mainly
    produced in low-Z MS binaries. }\label{fig:qdist_salpeter}
\end{figure}

\begin{figure}
    \centering
    \includegraphics[width=0.7\columnwidth]{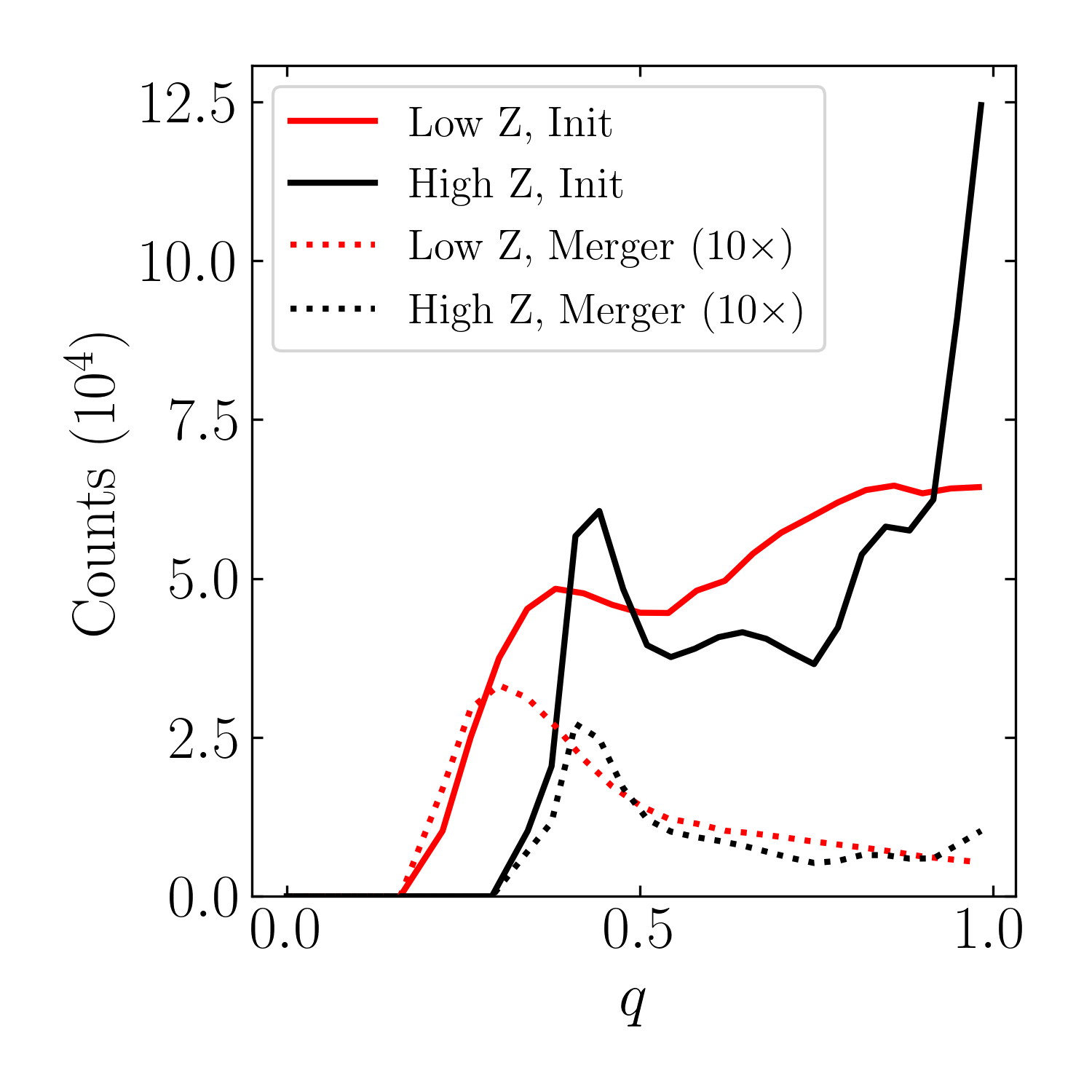}
    \includegraphics[width=\columnwidth]{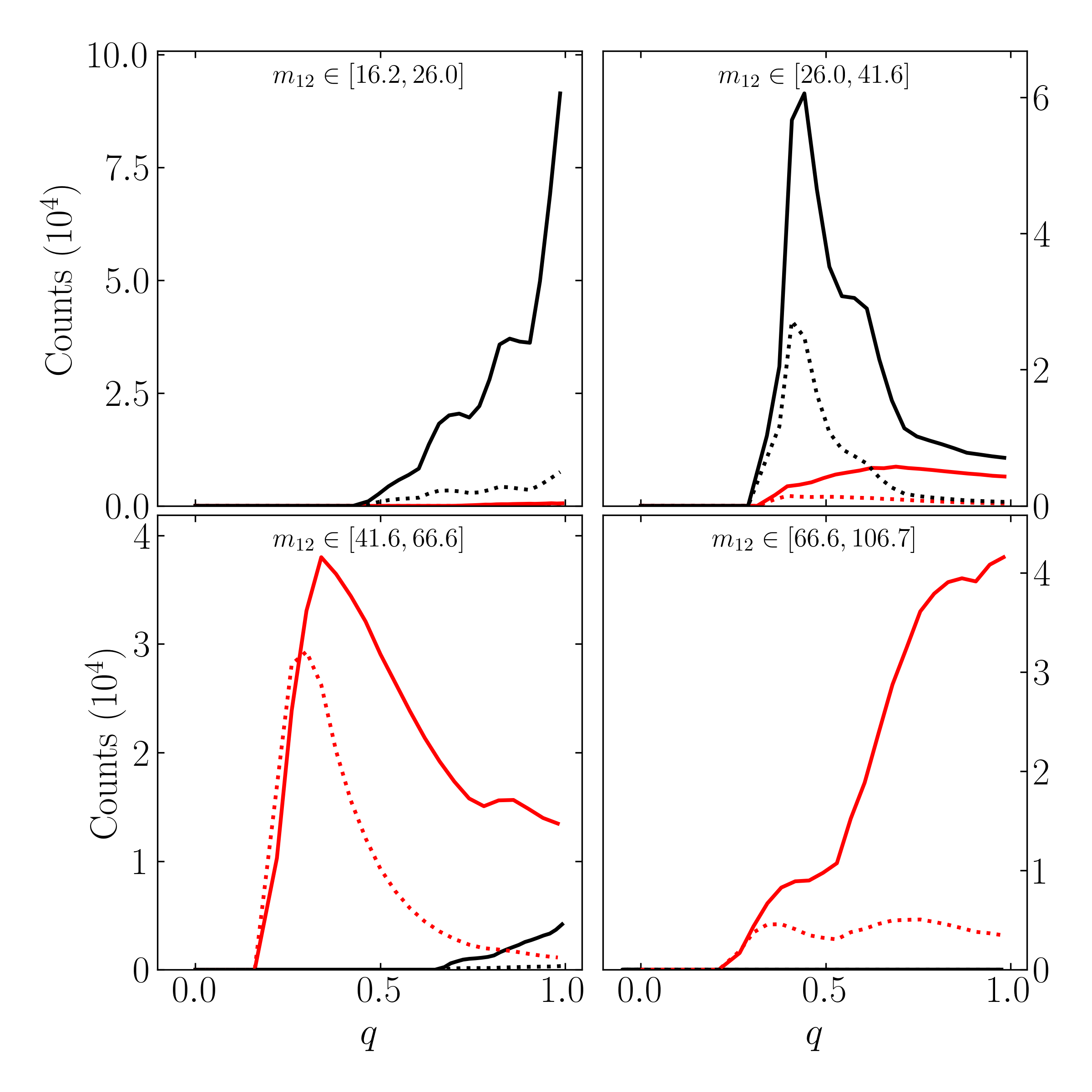}
    \caption{Same as Fig.~\ref{fig:qdist_salpeter} but for $\alpha = 2$, i.e.\ a
    nearly uniform distribution of the main sequence binary mass
    ratio. The results are very similar to Fig.~\ref{fig:qdist_salpeter}.
    }\label{fig:qdist_uniform}
\end{figure}

Using the results of Section~\ref{ss:qdist_init}, we can also estimate the mass
ratio distribution of merging BHs using Eq.~\eqref{eq:gmerge_dq}. We consider
representative triple systems considered in Section~\ref{s:merger_frac}: for
$\eta_{\rm merger}$, we use a simple approximation that lies roughly between the
two cases shown in Figs.~\ref{fig:popsynth}--\ref{fig:popsynth5500}:
\begin{align}
    \eta_{\rm merger}(q) \approx 0.2 \times \s{\max\p{q, 0.2}}^{-2}.
        \label{eq:etam_approx}
\end{align}
The results for $\rdil{F_{\rm merger}}{q}$ are displayed as the dotted curves in
Figs.~\ref{fig:qdist_salpeter}--\ref{fig:qdist_uniform} in each panel. Broadly
speaking, $\rdil{F_{\rm merger}}{q}$ peaks around $q \sim 0.3$ for low-Z
systems, and around $q\sim 0.4$ for high-Z systems, the latter reflecting the
peak in the initial BH binary q-distribution. Also note that $\rdil{F_{\rm
merger}}{q}$ can be quite different for different $m_{12}$ ranges. For example,
merging BH binaries with $m_{12} > 42M_\odot$ are only produced in low-Z
systems, and $\rdil{F_{\rm merger}}{q}$ peaks around $q\sim 0.3$ for $m_{12}\in
[42, 67]M_{\odot}$, and is roughly uniform between $q \sim 0.2$ to $1$ for
$m_{12}\gtrsim 67 M_{\odot}$.

We emphasize that these results for $\md F_{\rm merger}/\md q$ refer to the
representative triple systems studied in
Sections~\ref{s:background}--\ref{s:merger_frac}, and thus should be considered
for illustrative purposes only. As noted above, the merger fraction $\eta_{\rm
merger}$ depends on various parameters of the triple systems. While we have not
attempted to quantify $\eta_{\rm merger}$ for all possible triple system
parameters, it is clear that the principal finding of
Section~\ref{s:merger_frac} (i.e., $\eta_{\rm merger}$ increases with decreasing
$q$) applies only for systems with sufficiently strong octupole effects. In
fact, from Figs.~\ref{fig:total_merger_fracs} and~\ref{fig:sweepbin_simpleouter}
we can estimate that the octupole-induced feature in $\eta_{\rm merger}$ becomes
prominent only when $\epsilon_{\rm oct}\gtrsim 0.005$, or equivalently
\begin{equation}
  {a\over a_{\rm out,eff}} \gtrsim
    0.005 \left({1+q\over 1-q}\right){\sqrt{1-e_{\rm out}^2}\over e_{\rm out}}
\simeq {0.01 \over e_{\rm out}},
    \label{eq:new}
\end{equation}
where in the second step we have used $q\sim 0.5$ and $e_{\rm out}\sim 0.6$.
When this condition is satisfied, the inner binary can usually also undergo a
one-shot merger (see Eq.~\ref{eq:q_237}), leading to strong dependence of the
merger fraction on $q$. For triple systems with $a/a_{\rm out,eff}\lesssim 0.01$
(such as the case when the tertiary is a supermassive BH with $m_3\gtrsim 10^6
m_{12}$), the octupole effect is unimportant (see the discussion following
Eq.~\ref{eq:eps_oct}), and we expect the merger fraction to be almost
independent of $q$. Indeed, an analytical fitting formula for BH mergers induced
by pure quadrupole-ZLK effect shows $\eta_{\rm merger}\propto \mu^{0.16} \propto
q^{0.16}/(1+q)^{0.32}$ (see Eq.~53 of \citealp{LL18}, or Eq.~26 of
\citealp{bin_misc1}). For such systems, we expect $\md F_{\rm merger}/\md q$ to
be mainly determined by the initial $q$-distribution of BH binaries at their
formation.


\section{Summary and Discussion}\label{s:conclusion}

We have studied the dynamical formation of merging BH binaries induced by a
tertiary companion via the von Zeipel-Lidov-Kozai (ZLK) effect, focusing on the
expected mass ratio distribution of merging binaries. The octupole potential of
the tertiary, when sufficiently strong, can increase the inclination
window and probability of extreme eccentricity excitation, and thus enhance the
rate of successful binary mergers. Since the octupole strength $\epsilon_{\rm
oct} \propto (1-q)/(1+q)$ (see Eq.~\ref{eq:eps_oct}) increases with decreasing
binary mass ratio $q$, it is expected that ZLK-induced BH mergers favor binaries
with smaller mass ratios. We quantify the dependence of the merger
fraction/probability on $q$ using a combination of numerical integrations and
analytical calculations, based on the secular evolution equations for
hierarchical triples. We develop new analytical criteria
(Section~\ref{ss:nogw_merger}) that allow us to determine, without full
numerical integrations, whether an initial BH binary can undergo a ``one-shot
merger'' or a more gradual merger under the influence of a tertiary companion.
These allow us to compute the merger probability semi-analytically by only
studying non-dissipative (i.e.\ no GWs) triple systems (see
Eq.~\ref{eq:def_pmerge_sa}). We show that for hierarchical triples with
semi-major axis ratio $a/a_{\rm out}\gtrsim 0.01-0.02$ (see Eq.~\ref{eq:new}),
the BH binary merger fraction ($f_{\rm merger}$ or $\eta_{\rm merger}$) can
increase by a larger factor (up to $\sim 20$) as $q$ decreases from unity to
$0.2$ (see Figs.~\ref{fig:total_merger_fracs}--\ref{fig:popsynth_lowq}). When
combined with a reasonable estimate of the mass ratio distribution of the
initial BH binaries (Section~\ref{ss:qdist_init}), our results for the merger
fraction suggest that the final merging BH binaries have an overall mass ratio
distribution that peaks around $q = 0.3$ or $0.4$, although very different
distributions can be produced when restricting to specific ranges of total
binary masses (see Figs.~\ref{fig:qdist_salpeter} and~\ref{fig:qdist_uniform}).

Taking our final results (Figs.~\ref{fig:qdist_salpeter}
and~\ref{fig:qdist_uniform}) at face value, we tentatively conclude that the
mass-ratio distribution $\md F_{\rm merger}/\md q$ of BH binary mergers induced
by a comparable-mass companion is inconsistent with the current LIGO/VIRGO
result (see Fig.~\ref{fig:qhist}), suggesting that such tertiary-induced mergers
may not be the dominant formation channel for the majority of the detected
LIGO/VIRGO events. However, there are at least two important issues/caveats to
keep in mind:

(i) $\md F_{\rm merger}/\md q$ depends strongly on the initial mass-ratio
distribution of BH binaries at their formation ($\md F/\md q$), which is
uncertain and depends sensitively on the metalicity of the binary formation
environment (see Section~\ref{ss:qdist_init}). It is also possible that the
initial BH binary mass ratio distribution is much more skewed towards equal
masses than what we found in Section~\ref{ss:qdist_init} (e.g.\ if stellar
binaries with significantly asymmetric masses become unbound due to mass loss
and supernova kicks as their components become BHs). Such a
distribution was found by population synthesis studies that include
octupole-order ZLK effects and models of stellar evolution
\citep[e.g.][]{Hamers_q, toonen2018rate}. These studies find that ZLK
oscillations in stellar binaries with small $q$ can experience mass transfer and
merge without forming a compact object binary; as a result, most compact object
binaries form with large mass ratios. The prevalence of this phenomenon likely
depends on the initial semimajor axes of the inner binaries. Further study would
be required to understand the competition between this primordial large-$q$
enhancement and the elevated merger fractions for small $q$ found in the present
study in an astrophysically realistic population.

(ii) When the tertiary mass $m_3$ is much larger than the BH binary mass
$m_{12}$, as in the case of a supermassive BH tertiary, dynamical stability of
the triple requires $a_{\rm out} \gg a$, which implies that the octupole effect
is negligible ($\epsilon_{\rm oct}\ll 1$). For such triple systems, we expect
the merger fraction to depend very weakly on the mass ratio, and the final $\md
F_{\rm merger}/\md q$ to depend entirely on the initial $\md F/\md q$. Although
the merger fraction of such ``pure quadrupole'' triples is small ($\lesssim
6\%$; see Eq.~53 of \citealp{LL18}), additional ``external'' effects can enhance
the merger efficiency significantly [e.g., when the outer orbit experiences
quasi-periodic torques from the galactic potential
(\citealp{petrovich2017greatly}; see also \citealp{hamers2017secular}), or from
the spin of a supermassive BH \citep{LLW_apjl}].

Near the completion of this paper, we became aware of the simultaneous work by
\citet{martinez2021mass}, who study a similar topic using a population synthesis
approach.

\section{Acknowledgements}\label{s:ack}

We thank the anonymous referee whose detailed review and
comments greatly improved this paper. YS thanks Jiseon Min for useful
discussions. This work has been supported in part by NSF grant AST1715246. YS is
supported by the NASA FINESST grant 19-ASTRO19-0041. 
BL gratefully acknowledges support from the European Union's Horizon 2020
research and innovation program under the Marie Sklodowska-Curie grant agreement
No. 847523 `INTERACTIONS'.

\section{Data Availability}

The data referenced in this article will be shared upon reasonable request to
the corresponding author.

\bibliographystyle{mnras}
\bibliography{Su_EZLK}

\clearpage
\onecolumn

\appendix

\section{Origin of Octupole-Inactive Gap}\label{app:gap}

We investigate the origin of the ``octupole-inactive gap'', an
inclination range near $I_0 \approx 90^\circ$ for which $e_{\max}$ does not
attain $e_{\lim}$ despite being in between two octupole-active windows. This gap
was first identified in Section~\ref{ss:oct_gen}, and is seen in both the
non-dissipative and full simulations with GW dissipation (see
Figs.~\ref{fig:composite_dist}--\ref{fig:composite_bindist}).

\begin{figure}
    \centering
    \includegraphics[width=0.8\columnwidth]{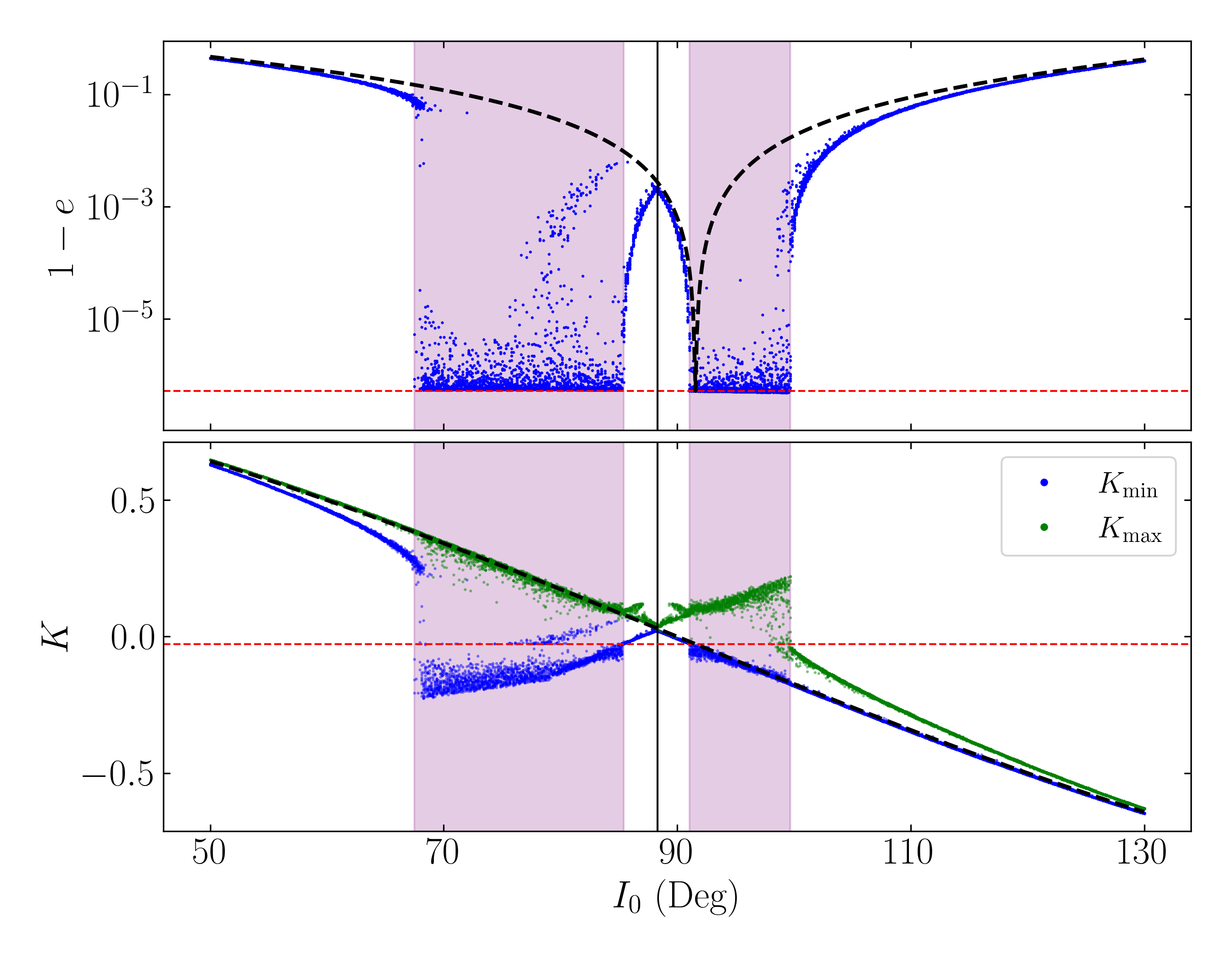}
    \caption{Octupole-active windows and amplitude of oscillation of $K$
    (Eq.~\ref{eq:def_K}). \emph{Top panel:} Maximum eccentricity $e_{\max}$
    attained by the inner binary with initial tertiary inclination $I_0$ when
    integrated for $2000 t_{\rm ZLK}$ (blue dots), reproduced from the top panel
    of Fig.~\ref{fig:composite_1p2}. Also shown are $e_{\lim}$
    (Eq.~\ref{eq:def_elim}, horizontal red dashed line), the quadrupole-level
    result for $e_{\max}$ (Eq.~\ref{eq:emax_quad}, dashed black line), the
    empirically-determined center of the gap, located at $I_0 \approx
    88.32^\circ$ (vertical black line), and the inclinations that can lead to
    extreme eccentricities (shaded purple regions). \emph{Bottom panel:} Minimum and maximum
    values of $K$, denoted $K_{\min}$ and $K_{\max}$, attained by the systems.
    Also shown are the initial $K$ for a given $I_0$ (black dashed line) and the
    critical $K_{\rm c} = -\eta / 2$ for orbit flipping (horizontal red dashed
    line). The center of the octupole-inactive gap and the octupole-active
    windows are labeled as in the top panel.
    }\label{fig:kdist}
\end{figure}

To better understand this gap, we first review the mechanism by which extreme
eccentricity excitation occurs. In the test-particle limit, \citet{katz2011long}
showed that $K$ (Eq.~\ref{eq:def_K}) oscillates over long timescales when
$\omega$, the argument of pericenter of the inner orbit, is circulating. This
then leads to orbit flips (and extreme eccentricity excitation) between prograde
and retrograde inclinations when $K$ changes signs: since $j(e)$ is nonnegative,
the sign of $K$ determines the sign of $\cos I$. \citet{katz2011long} obtained
coupled oscillation equations in $K$ and $\Omega_{\rm e}$, the azimuthal angle
of the inner eccentricity vector in the inertial reference frame. The amplitude
of oscillation of $K$ can then be analytically computed, and the octupole-active
window (the range of $I_0$ over which orbit flips occur) is the region for which
the range of these oscillations encompasses $K = 0$ \citep{katz2011long}.
When $\omega$ is librating instead, $\Omega_{\rm e}$ jumps by $\sim 180^\circ$
every ZLK cycle, and the oscillations in $K$ are suppressed.

In the finite-$\eta$ case, we commented in Section~\ref{ss:oct_gen} that the
relation between $K$ oscillations and extreme eccentricity excitation (and
orbit flipping) can be generalized even when $\eta$ is nonzero. $K$ still
oscillates over timescales $\gg t_{\rm ZLK}$ when $\omega$ is circulating, and
if its range of oscillation contains $K_{\rm c} \equiv -\eta / 2$, then the
inner orbit flips, in the process attaining extreme eccentricities. To be
precise, orbit flips are defined to be when the range of inclination
oscillations changes from $\p{\cos I_0}_- < \cos I < \cos I_{0, \lim}$ to $\cos
I_{0, \lim} < \cos I < \p{\cos I_0}_+$ or vice versa, where $\p{\cos I_0}_{\pm}$
are given by Eq.~\eqref{eq:I0bounds} and $I_{0, \lim}$ satisfies
Eq.~\eqref{eq:def_I0lim}.

However, the range of oscillation of $K$ is more complex than it is in the
test-particle limit. Figure~\ref{fig:kdist} compares the behavior of $e_{\max}$
in the non-dissipative simulations (top panel; reproduced from the top panel of
Fig.~\ref{fig:composite_1p2}) to the range of oscillations in $K$ (bottom
panel). Denote the center of the gap $I_{\rm 0, gap}$ (shown as the vertical
black line in both panels of Fig.~\ref{fig:kdist}). Near $I_{\rm 0, gap}$, $K$
oscillates about $K(I_{\rm 0, gap})$, which is \emph{positive}, and the
oscillation amplitude goes to zero at $I_{\rm 0, gap}$. On the other hand, orbit
flips (and extreme eccentricity excitation) are possible when the range of
oscillation of $K$ encloses $K_{\rm c}$ (i.e., $K_{\min} < K_{\rm c} <
K_{\max}$). The purple shaded regions in both panels of Fig.~\ref{fig:kdist}
illustrate this equivalence, as they show both the $e_{\lim}$-attaining
inclinations in the top panel and the inclinations where $K_{\min} < K_{\rm c} <
K_{\max}$ in the bottom panel. But since $K\p{I_{\rm 0, gap}} > 0$ while $K_{\rm
c} < 0$, there will always be a range of $I_0$ about $I_{\rm 0, gap}$ for which
the oscillation amplitude is smaller than $K\p{I_{\rm 0, gap}} - K_{\rm c}$, and
orbit flips are impossible in this range. This range then corresponds to the
octupole-inactive gap.

\begin{figure}
    \centering
    \includegraphics[width=0.45\columnwidth]{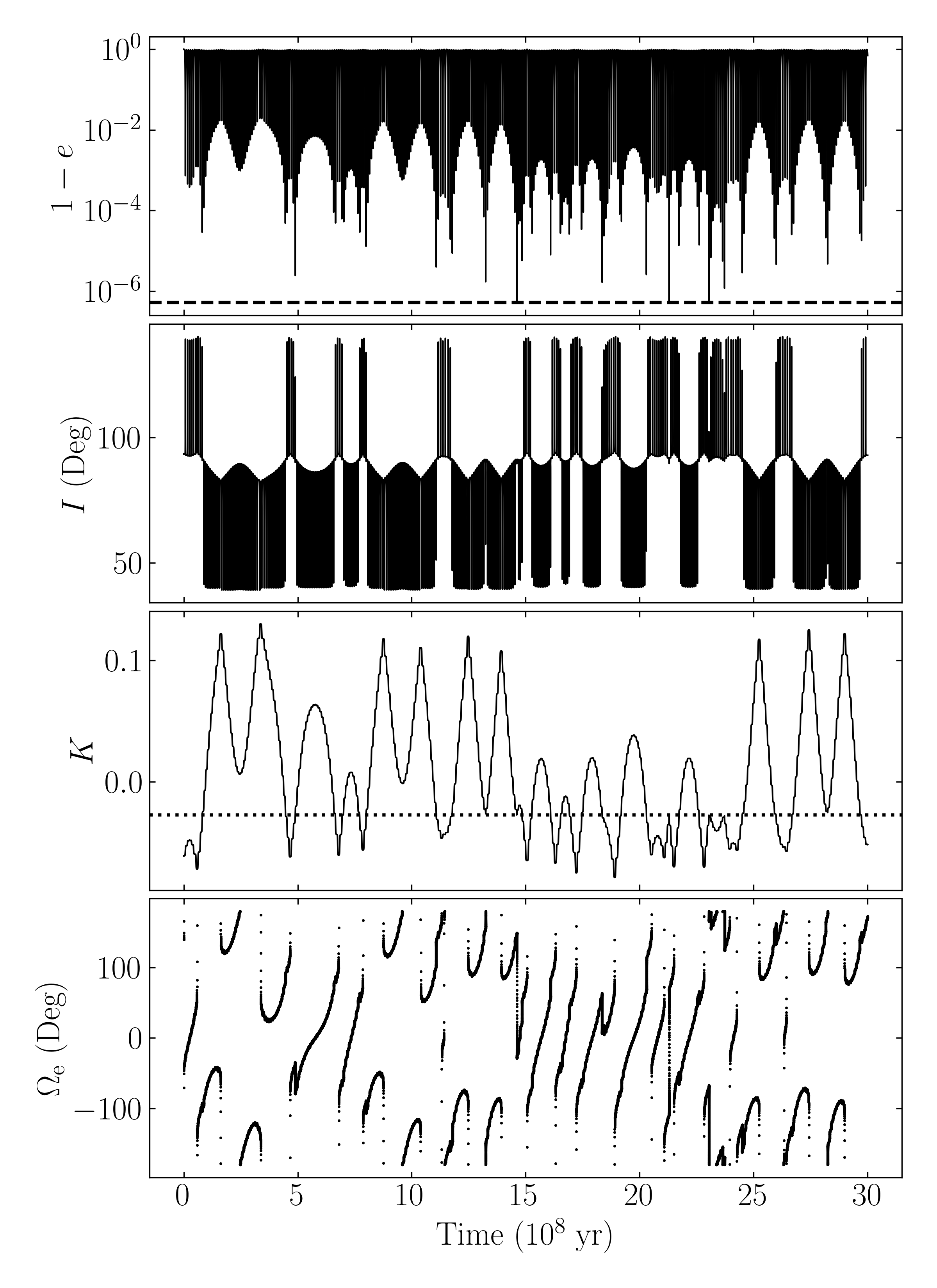}
    \includegraphics[width=0.45\columnwidth]{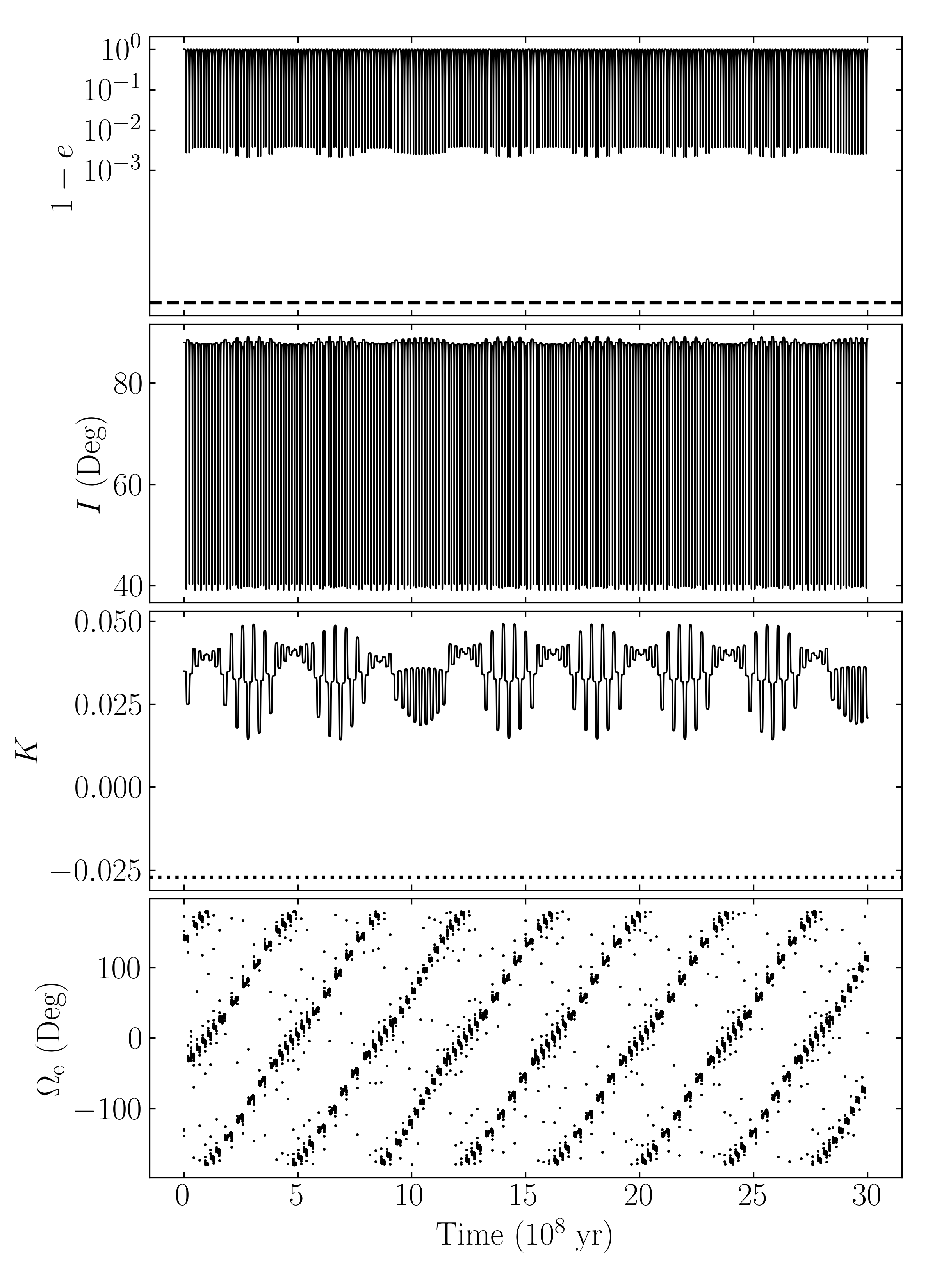}
    \caption{The left panel is the same as Fig.~\ref{fig:nogw_fiducial} but
    includes the evolution of the azimuthal angle of the eccentricity vector,
    $\Omega_{\rm e}$. The right panel is the same as the left but for $I_0 =
    88^\circ$. For both of these examples, we have used $\omega_0 = 0$, but the
    evolution is similar for all $\omega_0$.}\label{fig:nogw_circ}
\end{figure}

This analysis has simply pushed our lack of understanding onto a new quantity:
why are $K$ oscillations suppressed in the neighborhood of $I_{\rm 0, gap}$? A
quantitative answer to this question is beyond the scope of this paper, but for
a qualitative understanding, we can examine the evolution of a system in the
octupole-inactive gap. The left panel of Fig.~\ref{fig:nogw_circ} shows the same
simulation as Fig.~\ref{fig:nogw_fiducial} but with an additional panel showing
$\Omega_{\rm e}$, while the right panel shows a simulation with the same
parameters except $I_0 = 88^\circ$, which is near $I_{\rm 0, gap}$ (see
Fig.~\ref{fig:kdist}). The oscillations in $K$ (third panels) are much smaller
for $I_0 = 88^\circ$ than for $I_0 = 93.5^\circ$, and no orbit flips occur. Most
interestingly, the fourth panel shows that the evolution of $\Omega_{\rm e}$ is
much less smooth than in Fig.~\ref{fig:nogw_fiducial}, jumping at almost every
other eccentricity maximum. \citet{katz2011long} have already pointed out that
jumps in $\Omega_{\rm e}$ occur when $\omega$ is \emph{librating}, rather than
circulating.

\begin{figure}
    \centering
    \includegraphics[width=0.7\columnwidth]{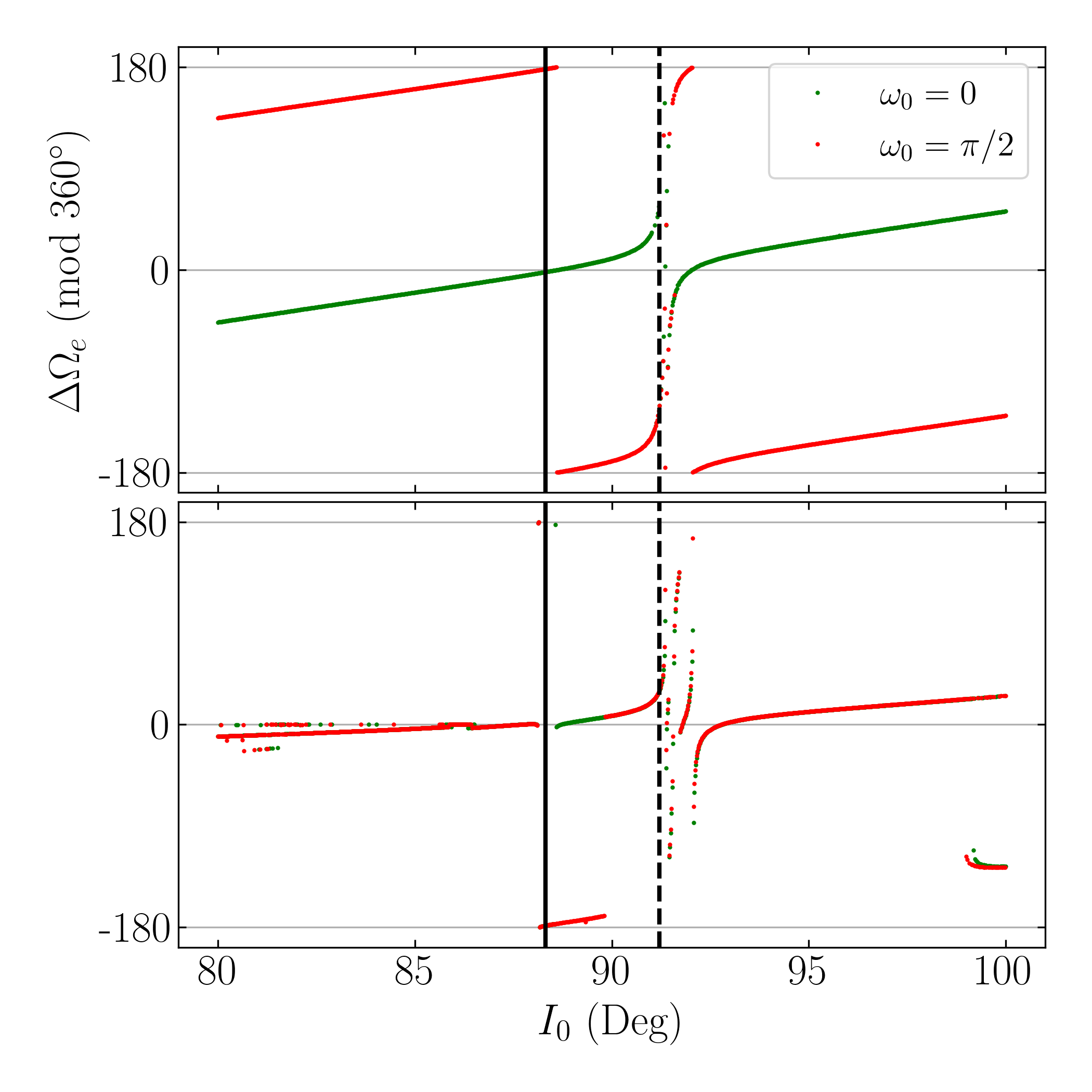}
    \caption{Plot of $\Delta \Omega_{\rm e}$, the change in $\Omega_{\rm e}$
    over a single ZLK cycle, for $q = 0.2$ and the fiducial parameters using
    different initial conditions. In the top panel, octupole-order terms are
    neglected, while in the lower panel, they are not. The solid and dashed
    vertical black lines denote $I_{\rm 0, gap}$ and $I_{0, \lim}$
    respectively.}\label{fig:dW}
\end{figure}

When the octupole-order terms are neglected, the circulation-libration
boundary is a boundary in $e$-$\omega$ space: as long as the ZLK separatrix
exists in the $e$-$\omega$ plane and $e_0 > 0$, then an initial $\omega_0 = 0$
causes $\omega$ to circulate, while an initial $\omega_0 = \pi/2$ causes
$\omega$ to librate \citep[e.g.,][]{kinoshita, shevchenko2016lidov}. However,
when including octupole-order terms, this picture breaks down. To illustrate
this, for a range of $I_0$ and both $\omega_0 = 0$ and $\omega_0 = \pi$, we
evolve the fiducial system parameters for a single ZLK cycle, using $q = 0.2$ as
is used for Figs.~\ref{fig:kdist} and~\ref{fig:nogw_circ}, and consider both the
dynamics with and without the octupole-order terms. Figure~\ref{fig:dW} gives
the resulting changes in $\Omega_{\rm e}$ over a single ZLK period when the
octupole-order effects are neglected (top) and when they are not (bottom). Two
observations can be made: (i) $I_{\rm 0, gap}$ is approximately where $\Delta
\Omega_{\rm e} = 0$ for circulating initial conditions when neglecting
octupole-order terms, and (ii) the inclusion of the octupole-order terms seem to
cause $\Omega_{\rm e}$ to exclusively vary slowly ($\abs{\Delta \Omega_{\rm e}}
\ll 180^\circ$) except for $I_{\rm 0, gap} < I_0 < I_{0, \lim}$. The former is
plausible: if $K(I_{\rm 0, gap})$ is the location of an equilibrium in
$K$-$\Omega_{\rm e}$ space, then it must satisfy $\Delta \Omega_{\rm e} = 0$.
The latter suggests that the assumption of circulation of $\omega$ in
\citet{katz2011long} may be satisfied for many more initial conditions than the
quadrupole-level analysis suggests, as long as they are not in octupole-inactive
gap.

Finally, examination of the bottom panel of Fig.~\ref{fig:kdist} suggests that
the oscillation amplitude in $K$ grows roughly linearly with $\abs{I_0 - I_{\rm
0, gap}}$ in the vicinity of $I_{\rm 0, gap}$ \citep[this may be because, in the
test-particle limit, librating $\omega$ give oscillation amplitudes in $K$ that
are higher-order in $K$ and $\Omega_{\rm e}$, as pointed out
by][]{katz2011long}. Assuming this, the gap width can then be given by
\begin{align}
    \text{Gap Width} = 2\p{I_{0, \lim} - I_{\rm 0, gap}}.
\end{align}
This explains why the gap does not exist in the test-particle regime, as $I_{0,
\lim} = I_{\rm 0, gap} = 90^\circ$ by symmetry of the equations of motion.

It is clear from the preceding discussion and Fig.~\ref{fig:dW} that the
octupole-order, finite-$\eta$ dynamics are complex, and our discussion can only
be considered heuristic. Nevertheless, in the absence of a closed form solution
to the octupole-order ZLK equations of motion or a full generalization of the
work of \citet{katz2011long}, they provide a preliminary understanding of the
octupole-inactive gap.

\label{lastpage} 
\end{document}